\documentclass[12pt]{article}

\AtBeginDocument{
	\addtocontents{toc}{\normalsize}
}
\pdfoutput=1

\usepackage{amsmath,amssymb,amsfonts,amsthm,amscd,mathrsfs}
\usepackage{esvect}
\usepackage{xcolor}
\definecolor{darkblue}{rgb}{0.1,0.1,.7}
\usepackage[]{version}
\usepackage[]{graphicx}
\usepackage[]{latexsym}
\usepackage{geometry}
\usepackage[all,cmtip]{xy}

\usepackage[margin=10pt,font=small,labelfont=bf]{caption}
\usepackage{ifthen}
\usepackage{soul}
\usepackage{tikz}
\usepackage{array,setspace,mathrsfs,amsfonts,yfonts,dsfont,bbm,colonequals}
\usepackage{dsfont}
\usepackage{cite}
\usepackage{xspace}
\usepackage{empheq}
\usepackage{extarrows}

\usepackage{xcolor}
\usepackage[colorlinks, linkcolor=darkblue, citecolor=darkblue, urlcolor=darkblue, linktocpage]{hyperref} 

\usepackage{subcaption}
\usepackage[utf8]{inputenc}
\usepackage{setspace}
\usepackage{float, graphicx}

\numberwithin{equation}{section}

\newcommand{\cO}{\mathcal O}

\newcommand{\reef}[1]{(\ref{#1})}
\newcommand{\be}{\begin{equation}}
\newcommand{\ee}{\end{equation}}
\newcommand{\bea}{\begin{eqnarray}}
\newcommand{\eea}{\end{eqnarray}}
\newcommand{\ba}{\begin{equation}\begin{aligned}}
\newcommand{\ea}{\end{aligned}\end{equation}}

\newcommand{\ud}{\mathrm d}
\newcommand{\mbf}{\mathbf}

\newcommand{\Df}{{\Delta_\phi}}

\def\s{\sigma}

\def\a{\alpha}
\def\e{\epsilon}
\def\k{\kappa}
\def\b{\beta}
\def\d{\delta}

\def\D{\Delta}

\newcommand{\mf}[1]{\mathfrak #1}

\begin{document}
	
	\vspace*{-.6in} \thispagestyle{empty}
	\begin{flushright}
		LPTENS/18/18
	\end{flushright}
	\vspace{1cm} {\Large
		\begin{center}
			{\bf Charging Up the Functional Bootstrap}\\
	\end{center}}
	\vspace{1cm}
	\begin{center}
		{\bf Kausik Ghosh${}^{S} $, Apratim Kaviraj${}^{T,A} $ , Miguel F.~Paulos${}^{A}$ }\\[1cm] 
		{
			\small
			{\em  ${}^{S} $Centre for High Energy Physics,
				 Indian Institute of Science,\\  C.V. Raman Avenue, Bangalore 560012, India. \\ ${}^T$DESY Hamburg, Theory Group, Notkestra\ss e 85, D-22607 Hamburg, Germany,\\
				 Institut de Physique Th\'{e}orique Philippe Meyer \&  \\
				 ${}^A$Laboratoire de Physique Th\'eorique, \\ de l'\'Ecole Normale Sup\'erieure, PSL University,\\ CNRS, Sorbonne Universit\'es, UPMC Univ. Paris 06\\ 24 rue Lhomond, 75231 Paris Cedex 05, France
			}
			\normalsize
		}
		\\
	\end{center}
	
	\begin{center}
		{\texttt{kau.rock91@gmail.com,} \ \ \texttt{apratim.kaviraj@desy.de,} \ \  \texttt{miguel.paulos@ens.fr} 
		}
		\\
	\end{center}
	
	\vspace{4mm}
	
	\begin{abstract}
		We revisit the problem of bootstrapping CFT correlators of charged fields. After discussing in detail how bounds for uncharged fields can be recycled to the charged case, we introduce two sets of analytic functional bases for correlators on the line. The first, which we call ``simple'', is essentially a direct sum of analytic functionals for the uncharged case. We use it to establish very general bounds on the OPE density appearing in charged correlators. The second basis is dual to generalized free fields and we explain how it is related to a charged version of the Polyakov bootstrap. We apply these functionals to map out the space of correlators and obtain new improved bounds on the 3d Ising twist defect.
	\end{abstract}
	\vspace{2in}

	
	\newpage
	
	{
		\setlength{\parskip}{0.05in}
		\tableofcontents
		\renewcommand{\baselinestretch}{1.0}\normalsize
	}
	
	
	\setlength{\parskip}{0.1in}
	\newpage

	\section{Introduction}
	Crossing equations in conformal field theories (CFTs) are remarkable objects: while we believe they encode almost if not all information about almost if not all critical phenomena, cajoling these equations into yielding useful information is not an  easy task. For this reason they have fascinated humankind since the dawn of time. 
	
	The typical bootstrapper's tool for analysing these equations is brute-force Taylor expansion. This turns out to be remarkably effective and has taken us quite far \cite{Poland:2018epd}, but this does not mean we cannot do better. A subtler approach has been proposed based on the concept of analytic `extremal functionals’ \cite{El-Showk:2016mxr, Mazac:2016qev, Mazac:2018mdx,Mazac:2018ycv} (see also \cite{Mazac:2018qmi,Mazac:2019shk,Caron-Huot:2020adz, Sinha:2020win, Gopakumar:2021dvg, Penedones:2019tng, Carmi:2020ekr,  Kaviraj:2018tfd,Mazac:2018biw,Giombi:2020xah} for related works). Such functionals are dual to extremal solutions to crossing, i.e. one which saturates numerical bootstrap bounds. The functionals thus naturally lead to optimal bounds on the CFT data. This approach is intimately related to the so-called Polyakov Bootstrap \cite{Polyakov}, wherein one  writes a CFT correlator in a way which manifests crossing but obscures the OPE \cite{Gopakumar:2016cpb, Gopakumar:2016wkt, Gopakumar:2018xqi} (see also \cite{Sen:2015doa}).

	In this paper we take a modest step in our quest for understanding arbitrarily complicated systems of crossing equations, by extending the analytic functional bootstrap to tackle CFT correlators of fields charged under a global symmetry \cite{Kos:2013tga, Kos:2015mba}. The functional bases of \cite{Mazac:2019shk,Caron-Huot:2020adz} for higher dimensional CFTs can be trivially extended to this setup as we will mention in the discussion, essentially because they do not care about full crossing symmetry. Here we will focus on 1d CFTs, or more precisely, to CFT correlators restricted to the line, where it is indeed possible to have full crossing symmetry.

	There are several motivations for doing this. Firstly, there are interesting physical systems described by 1d CFTs with global symmetries, such as long-range versions of familiar statistical physics models such as the $O(N)$ models. These can be thought of as toy models for higher dimensional systems which nevertheless provide highly non-trivial examples of families of conformal fixed points. Another large class of examples consists of conformal line defects, where rotations in the embedding spacetime descend to a global symmetry on the defect. More trivially, restricting any higher-dimensional CFT on a line also leads a 1d conformal system with a global symmetry. Finally, global symmetry provides us with a system of multiple bootstrap equations, and setting up analytic functionals for this case is the first step towards to do the same for multiple correlators involving distinct conformal primaries.

	 We will show that for a wide class of charged CFT correlators we can construct two types of analytic functional bases. The first type, which we call ‘simple’ is a set of functionals that is essentially given as a direct sum of functionals for the uncharged case. The second type of basis, ‘GFF functionals’ is a functional basis that is dual to the Generalized Free Field solution for the specific group we considered. We show this basis of functionals can also be obtained from a version of the Polyakov bootstrap for charged correlators, clarifying and rigorously justifying some of the statements in \cite{Ferrero:2019luz}. We show that this basis can be encoded into a `master functional', whose sum rules translates into a crossing-symmetric dispersion relation for charged CFT correlators.
	
	Both kinds of functionals are useful in applications. We use the simple functional basis to obtain bounds on OPE coefficients in various charged channels. In particular we demonstrate that in any such channel, the OPE density is bounded from above by that in a GFF-type solution up to an order one coefficient. We then use the GFF functional basis to demonstrate two numerical applications. The first one is to map out the space of allowed correlator values for the $O(N)$ case. We find this space has several corners, some of which we understand analytically, others which we don't and could be candidates for interesting CFTs. The second application is to revisit the bootstrap of the 3d Ising twist defect, introduced in \cite{Billo:2013jda} and first bootstrapped in \cite{Gaiotto:2013nva}. We find that the analytic functional basis, in this case for the fundamental irrep of the $O(2)$ group, leads to dramatic improvements in the convergence of the numerics, leading to significantly improved predictions for the twist defect spectrum. 
	
	The plan of this paper is as follows: in section \ref{sec:kinematics} we discuss the kinematics for describing and bootstrapping a large class of charged correlators. In section \ref{sec:simplefuncs} we define the simple functional basis and prove general bound for the charged OPE densities appearing in such correlators. We describe the Polyakov bootstrap in section \ref{sec:PolyakovBs} and propose a generalization for the charged case. This proposal is justified in section \ref{sec:GFFfunctionals} by the construction of a new set of `GFF functionals', which we also show to be captured by a crossing-symmetric dispersion relation for charged correlators. In section \ref{sec:numerics} we consider two numerical applications of GFF functionals, first mapping out the space of allowed $O(N)$ correlators, and then focusing on $O(2)$ to bootstrap the Ising twist defect. We conclude in section \ref{sec:conclusion} with a short discussion and outlook. The paper is completed by several technical appendices.
	
	\section{Kinematics}
	\label{sec:kinematics}
	We are interested in correlators of fields transforming in a unitary real irreducible representation (irrep) $\mf r$ of dimension $d_{\mf r}$ of a compact symmetry group\footnote{So this means the action of the group can always be represented by real orthogonal matrices}. In this work we will consider cases where the tensor product $\mf r \otimes \mf r$ decomposes into a sum of $r$ independent real irreps $\mf a$ appearing with {\em unit} multiplicity.\footnote{If this was not the case, the OPE decomposition of the correlator would include non-positive products of OPE coefficients.} Note that one of these irreps is necessarily the trivial one, which we call singlet and will indicate by $\mf a=\mbf S$.
	
	Let us denote the field by $\phi_i$, where $i=1,\ldots,d_{\mf r}$. The four point function restricted to the line takes the form
	\begin{align}\label{4ptdef}
	\langle \phi_i (x_1) \phi_j (x_2) \phi_k (x_3)
	\phi_l (x_4) \rangle = \frac{\mathcal{G}_{ijkl}(z)}{(x_{13}^2)^{\Delta_{\phi}}
		(x_{24}^2)^{\Delta_{\phi}}}, \hspace{0.5cm}z^2 = \frac{x_{12}^2 x_{34}^2}{x_{13}^2 x_{24}^2}, \hspace{0.5cm}x_{ij}^2=(x_i-x_j)^2.
	\end{align}
	and we have the crossing relation
	\bea
	\mathcal G_{ijkl}(z)=\mathcal G_{ilkj}(1-z)
	\eea
	We can now introduce a basis of invariant tensor structures $T^{\mf a}_{ijkl}$ associated to each exchanged irrep $\mf a$ in the $\phi^{\mf r} \times \phi^{\mf r}$ OPE. These can be expressed in terms of Clebsch-Gordan coefficients:
	\ba
	T^{\mf a}_{ij,kl}=\frac{1}{\sqrt{d_{\mf a}}}\sum_{s=1}^{d_{\mf a}}  (\mathcal C^{\mf a})_{ij,s}  (\mathcal C^{\mf a})_{ lk}^{\ \ s}\,,\qquad \sum_{i,j=1}^{d_{\mf r}}\,  (\mathcal C^{\mf a})_{ij,s} (\mathcal C^{\mf b})^{ij,r}=\delta_{s}^r \delta^{\mf a \mf b}
	\ea
	with $d_{\mf a}$ the dimensionality of the irrep $\mf a$. While in general there can be other four-point invariant tensor structures for a given group, the above are the only ones relevant for decomposing a CFT correlator consistent with an OPE. These tensor structures satisfy the relations
	\ba
	T^{\mf a}_{ij,kl}&= \sum_{\mf b} C^{\mf b \mf a} T^{\mf b}_{il,kj}\,,& \qquad T^{\mf a}_{ij,kl}&=\sum_{\mf b} \eta^{\mf b \mf a} T^{\mf b}_{ji,kl}\,,\\
	T_{ij,kl}^{\mf a}&=T_{kl,ij}^{\mf a}\,,& \qquad \mathcal \delta^{\mf a \mf b}&=\sum_{ijkl} T^{\mf a}_{ij,kl}(T^{\mf b})^{ij,kl}
	\ea
	where indices are raised/lowered with the invariant tensor $\delta^{i}_j$.
	Note that among these the tensor structure associated to the singlet is special:
	\ba
	T^{\mbf S}_{ij,kl}=\frac{1}{d_{\mf r}} \delta_{ij}\delta_{kl}\,.
	\ea
	From these definitions it easily follows:
	\ba
	C=C^{T}\,,\qquad C\cdot C=\eta\cdot\eta=1\,, \qquad C\cdot \eta\cdot C=\eta\cdot C\cdot \eta
	\ea
	The $\eta$ matrix is diagonal, and we have: 
	\bea
	\eta^{\mf a \mf b}=\delta^{\mf a \mf b} \eta^{\mf a}\,, \qquad \eta^\mf a=\pm 1\,, \qquad \eta^{\mbf S}=1\,.
	\eea
	We say an irrep has even/odd parity if $\eta^{\mf a}=+1(-1)$ respectively. In what follows we will assume without loss of generality the identity:
	\bea
	T^{\mf a}_{[ij,k]l}=0 \quad \Rightarrow \quad (1-C+\eta\cdot C)\cdot (1-\eta)=0 \label{eq:parityoddid}
	\eea
	The reason for this is that if there is some structure that is nonzero under this antisymmetry property then it is crossing antisymmetric as well as parity odd by itself. It therefore decouples from the more interesting irreps satisfying the above which in general do mix under crossing. As will become clear below, contributions to the correlator associated to such irreps could be bootstrapped independently as in a problem without global symmetry.
	
	It is useful to introduce projectors onto crossing (anti)symmetric structures:
	\bea
	P_{\pm}^{\mf a \mf b}:= \frac 12 \left( \delta^{\mf a \mf b}\pm C^{\mf a \mf b}\right)\,.
	\eea
	Since $P_{\pm}^{\mf a \mf b}$ are projectors we can write them as
	\bea
	P_{\pm}^{\mf a \mf b}=\sum_{s=1}^{r_{\pm}} e_{\pm,s}^{\mf a} e_{\pm,s}^{\mf b}\,, \label{eq:proj}
	\eea
	where $r_++r_-=r$ is the number of independent irreps in $\mf r\otimes \mf r$ and
	\bea
	\sum_{\mf b}  e_{\pm,s}^{\mf b} e_{\pm,y}^{\mf b}=\delta_{st}\,, \qquad \sum_{\mf b}  e_{\pm,s}^{\mf b} e_{\mp,t}^{\mf b}=0\,.
	\eea
	The quantities $e^{\mf a}_{\pm,s}$ are eigenvectors of the crossing matrix $C^{\mf a \mf b}$,
	\bea
	\sum_{\mf b} C^{\mf a \mf b} e_{\pm, s}^{\mf b}= \sum_{\mf b}  e_{\pm, s}^{\mf b}C^{\mf b \mf a} =\pm e_{\pm, s}^{\mf a}\,.
	\eea
	At this point we can remark that the numbers of $\pm 1$ eigenvectors are the same for both $C$ and $\eta$, since
	\bea
	r_+-r_-=\mbox{Tr}( C )=\mbox{Tr}(\eta\cdot C \cdot \eta)=\mbox{Tr}(C \cdot \eta \cdot C)=\mbox{Tr}(\eta).
	\eea

	We can use all this technology to express the correlator, as well as the crossing equation, in a nicer form:
	\bea
	\mathcal G_{ijkl}(z)=\sum_{\mf a} T^{\mf a}_{ijkl} \mathcal G^{\mf a}(z)\,, \qquad \mathcal G^{\mf a}(z)=\sum_{\mf b} C^{\mf a \mf b} \mathcal G^{\mf b}(1-z)\,.
	\eea
	Each $\mathcal G^{\mf a}(z)$ captures the contribution of a given representation $\mf a$ exchanged in the OPE. In fact the OPE gives
	\bea
	\mathcal G^{\mf a}(z)=\sum_{\Delta} a^{\mf a}_{\Delta} G_{\Delta}(z|\Df)
	\eea
	where the sum runs over $SL(2,\mathbb{R})$ primary operators in the $\phi \times \phi$ OPE, with OPE coefficient squared $\lambda_{\phi \phi \cO_\Delta^{\mf a}}^2\equiv a^{\mf a}_{\Delta}$ and $G_\D$ an $SL(2,\mathbb{R})$ conformal block, given by:
	\be
	G_{\D}(z|\Df)=z^{\D-2\D_\phi}{}_2F_1(\D,\D,2\D,z)\,.
	\ee
	Let us introduce crossing (anti)-symmetric `vectors':
	\bea
	F_{\pm,\Delta}(z|\Df)=G_{\D}(z|\Df)\pm G_{\D}(1-z|\Df)\,.
	\eea
	In terms of these the crossing equations become (suppressing $\D_\phi$ dependence for simplicity)
	\bea
	\sum_{\mf b}\sum_{\Delta} a^{\mf b}_{\Delta} \mathcal F^{\mf b|\mf a}_{\Delta}(z)=0\,, \qquad \mathcal F^{\mf b|\mf a}_{\Delta}(z)=P_{+}^{\mf a\mf b} F_{-,\Delta}(z)+P_-^{\mf a \mf b}\, F_{+,\Delta}(z)\,.\label{eq:crossingeq}
	\eea
		\begin{figure}[t]
		\begin{center}
			\includegraphics[width=7.5cm]{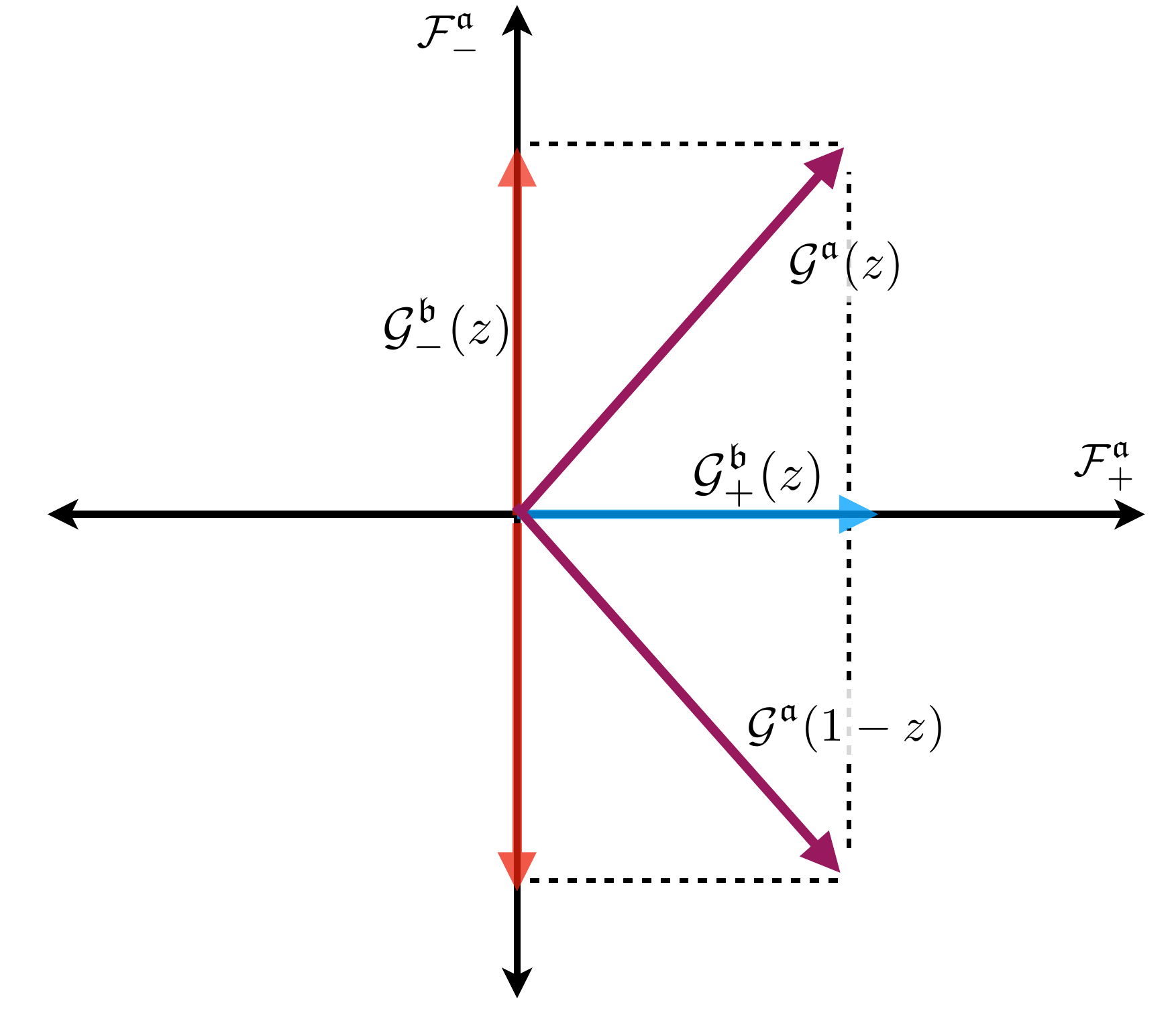}
			\caption{\label{fig:projections} Projections of $\mathcal{G}^{\mf a}(z)$ (violet) written as $\mathcal{G}_{\pm,s}(z)=\sum_{\mf a}e^{\mf{a}}_{\pm,s}\mathcal{G}^{\mf a}(z)$ sepatate out as crossing symmetric (blue) and crossing anti-symmetric vectors (red). }
		\end{center}
	\end{figure}
  It can be useful to diagonalize the crossing equation by introducing a concrete basis for the crossing-(anti)symmetric subspaces in the space of irreps (see figure \ref{fig:projections}). This is achieved by writing out the projectors as in \reef{eq:proj}. For instance we can write
	\ba
	\mathcal G^{\mf a}(z)&=\sum_{\mf b}\left[ P_{+}^{\mf a \mf b} \mathcal G^{\mf b}(z)+ P_{-}^{\mf a \mf b} \mathcal G^{\mf b}(z)\right]\\
	&=\sum_{s=1}^{r_+}  e_{+,s}^\mf a\mathcal G_{+,s}(z)+\sum_{s=1}^{r_-} e_{-,s}^{\mf a}\mathcal G_{-,s}(z)
	\ea
	where
	\bea
	\sum_{\mf a} e_{\pm,s}^{\mf a}\mathcal G^{\mf a}(z)=\mathcal G_{\pm,s}(z)=\pm \mathcal G_{\pm,s}(1-z).
	\eea
	The crossing equation for each such function is obtained by contracting 
	\reef{eq:crossingeq} with the eigenvectors $e_{\pm,\mu}^{\mf a}$ to get
	\ba
	\sum_{\Delta} a_{\pm,s,\Delta} F_{\mp,\Delta}(z)=0\,, \qquad a_{\pm,s,\Delta}=\sum_{\mf a} e^{\mf a}_{\pm,s} a^{\mf a}_{\Delta} \label{eq:decoupledcrossing}
	\ea
	
	This makes it clear that the crossing equation can be split into the analysis of $r$ seemingly decoupled functions, $r_+$ of which are crossing symmetric and $r_-$ crossing antisymmetric. In reality, these functions are linked by unitarity of $\mathcal G^{\mf a}$, translated into positivity of the OPE density $a^{\mf a}_{\Delta}$. This then constraints the OPE decompositions of these otherwise decoupled functions.
	
	\paragraph{General examples}
	
	It is useful to consider general examples of correlators which are both unitary and crossing symmetric. Consider the vector:
	\ba
	E_{+}^{\mf a}:=P_{+}^{\mf a \mbf S}=\frac 12\left(\delta^{\mf a\mbf S}+\frac{\sqrt{d_{\mf a}}}{d_{\mf r}}\right) \label{eq:allposguy}
	\ea
	This is an eigenvector of $C^{\mf a \mf b}$ with strictly positive components. Therefore it is always possible to choose a basis for the linear space spanned by the $+1$ eigenvectors of the crossing matrix, such that all basis elements have non-negative  components. Denoting such a basis by $E_{+,m}^{\mf a}$ then for any set of $r_+$ unitary solutions to the crossing symmetry problem {\em without} global symmetry, $\mathcal G_{+,k}(z)$, we have that
	\bea
	\mathcal G^{\mf a}(z)=\sum_{i=1}^{r_+} E_{+,i}^{\mf a} \mathcal G_k(z) \label{eq:sumsolution}
	\eea
	is also unitary and crossing symmetric. Note that a similar construction does not work in the anti-crossing symmetric sector: if we could make a $-1$ eigenvector of the crossing matrix with all positive components, it would still have to be orthogonal to all the $+1$ eigenvectors, including the vector $E_+^{\mf a}$ above. But this is impossible since they would have positive overlap.
	
	A simple but important example of a correlator satisfying crossing symmetry corresponds to setting $\phi_i$ to be a generalized free field, which in 1d we can take to be a boson or fermion. In this case we have
	\bea
	\mathcal G_{\mbox{\tiny free}}^{\mf a}(z)=\frac{\delta^{\mf a \mbf S}}{z^{2\Df}}+\frac{C^{\mf a \mbf S}}{(1-z)^{2\Df}}+\epsilon\, \eta^{\mf a}C^{\mf a \mbf S}\,. \label{eq:gfree}
	\eea
	with $\epsilon=+1$ or $-1$ sign for a bosonic or fermionic field respectively. This example gives for the OPE
	\bea
	\mathcal G^{\mf a}_{\mbox{\tiny free}}=\frac{\delta^{\mf a \mbf S}}{z^{2\Df}}+\sum_{n=0}^{\infty} C^{\mf a \mbf S} a^{\mbox{\tiny free}}_{\Delta_n^{\mf a}} G_{\Delta_n^{\mf a}}(z|\Df)
	\eea
	where
	\bea
	a^{\mbox{\tiny free}}_{\Delta}=\frac{2 \Gamma (\Delta)^2}{\Gamma (2 \Delta - 1)} \frac{\Gamma (\Delta + 2
		\Delta_{\phi} - 1)}{\Gamma (2 \Delta_{\phi})^2 \Gamma (\Delta - 2
		\Delta_{\phi} + 1)}\,. \label{eq:opefree}
	\eea
	As for the scaling dimensions they depend on whether we are considering the fermionic (F) or bosonic (B) cases. We have
	\ba
	\epsilon\, \eta^{\mf a}&=+1&\quad \Rightarrow \quad \Delta_n^{\mf a}&=2\Df+2n\equiv \Delta_n^B\\
	\epsilon\, \eta^{\mf a}&=-1&\quad \Rightarrow \quad \Delta_n^{\mf a}&=2\Df+2n+1\equiv \Delta_n^F\,.\label{eq:dimsgff}
	\ea
	In particular notice that the spectrum depends also on the parity of the exchanged irrep. It is then easy to see that this solution cannot be represented in the form \reef{eq:sumsolution}.
	
	\paragraph{O(N) application}
	As an illustration of the formalism let us consider the case where the field $\phi$ transforms in the fundamental representation $\mbf N$ of $O(N)$. In this case the tensor product contains representations which we label $\mbf S,\mbf T, \mbf A$, respectively the singlet, traceless symmetric, and antisymmetric representations, with dimensions $d_{\mbf S}=1, d_{\mbf T}= N(N+1)/2-1$ and $d_{\mbf A}=N(N-1)/2$. In this case we have
	\bea
	T^{\mbf S}_{ij,kl}=\frac {\d_{ij}\d_{kl}}{N}\,, \ T_{ijkl}^{\mbf T}=\frac 1{2\sqrt{d_{\mbf T}}} (\d_{ik}\d_{jl}+\d_{il}\d_{jk}-\frac{2}{N}\d_{ij}\d_{kl})\,,\ T_{ij,kl}^{\mbf A}=\frac { (\d_{il}\d_{jk}-\d_{ik}\d_{jl})}{2\sqrt{d_{\mbf A}}}\,.
	\eea
	The crossing matrix is given by
	\bea
	C^{\mf a \mf b}=\frac{1}{2N}
	\left(
	\begin{array}{ccc}
		2 & 2\sqrt{d_{\mbf T}} & 2\sqrt{d_{\mbf A}} \\
		2\sqrt{d_{\mbf T}} & N-2 & -N \sqrt{\frac{d_{\mbf T}}{d_{\mbf A}}} \\
		2\sqrt{d_{\mbf A}} &  -N \sqrt{\frac{d_{\mbf T}}{d_{\mbf A}}}  & N \\
	\end{array}
	\right)^{\mf a \mf b}\,, \qquad \mf a, \mf b \in \{\mbf S, \mbf T, \mbf A\}
	\eea
	The GFF solution is customarily written as
	\bea
	\mathcal G_{ijkl}(z)=\frac{\delta_{ij} \delta_{kl}}{z^{2\Df}}+\frac{\delta_{il} \delta_{jk}}{(1-z)^{2\Df}}+\epsilon \delta_{ik}\delta_{jl}\,.
	\eea
	This translates into
	\ba
	\mathcal G^{\mbf S}(z)&=\frac{N}{z^{2\Df}}+\frac{1}{(1-z)^{2\Df}}+\epsilon\\
	\mathcal G^{\mbf T}(z)/\sqrt{d_{\mbf T}}&=\frac{1}{(1-z)^{2\Df}}+\epsilon\\
	\mathcal G^{\mbf A}(z)/\sqrt{d_{\mbf A}}&=\frac{1}{(1-z)^{2\Df}}-\epsilon
	\ea
	This agrees perfectly with \reef{eq:gfree} up to an overall factor of $N$ related to our convention for normalizing the identity operator contribution.
	
	It is also interesting to consider the projections onto crossing-(anti)symmetric subspaces. Consider the following eigenvector basis for the crossing matrix:
	\ba
	E_{+,1}^{\mf a}&=\left(1,\frac{N-1}{\sqrt{d_{\mbf T}}},0\right)\,,& \quad  E_{+,2}^{\mf a}&=\left(1,0,\frac{N-1}{\sqrt{d_{\mbf A}}}\right)\,,& \quad E_{-,1}^{\mf a}&=\left(-2,\frac{2+N}{\sqrt{d_{\mbf T}}},\frac{N}{\sqrt{d_{\mbf A}}}\right)
	\ea
	As promised, the coefficients in $E_{+,1}, E_{+,2}$ could be chosen  positive. As discussed above, a simple set of unitary and crossing symmetric correlators can now be obtained by choosing any two unitary solutions to the uncharged crossing equation and setting
	\ba
	\mathcal G^{\mf a}(z)\propto E_{+,1}^{\mf a} \mathcal G_1(z)+E_{+,2}^{\mf a} \mathcal G_2(z)
	\ea
	with the proportionality constant fixed by demanding the right coefficient for the identity operator.

	\section{Simple functionals and their bounds}
	\label{sec:simplefuncs}
	In this section we discuss a particularly simple set of analytic functionals which act on the charged crossing equation. They are obtained simply as a direct sum of functionals which act on the {\em uncharged} equation. This construction allows us to promote known analytic functional bases, and their associated bounds, from the uncharged case to the present charged setup.
	
	\subsection{Review: uncharged case}
	
	Let us begin by discussing crossing equations without global symmetry. They take the form:
	\ba
	\sum_{\Delta} a_{\Delta} F_{\pm,\Delta}(z)=0\,.
	\ea
	The usual 1d CFT uncharged crossing equation is of $-$ type, but as we've seen in the previous section, with global symmetry we also get equations involving $F_{+,\Delta}$.
	
	A general approach for extracting constraints out of this equation is to introduce linear functionals, satisfying the property that they should commute over the infinite sum above~\cite{Qiao:2017lkv}:
	\be
	\omega_{\pm} \left[\sum_\D a_{\D}F_{\pm,\D}(z)\right]=\sum_\D a_{\D}\omega_{\pm} (\D)=0\,.
	\ee
	Here we set $\omega_{\pm} (\D)\equiv \omega_{\pm}[F_{\pm,\D}(z)]$, with the $\D_\phi$ dependence left implicit. Also, it is understood that a `$-$' type functional only acts on antisymmetric functions and conversely for a `$+$' type functional.
	
	In the usual numerical bootstrap approach, a useful complete set of functionals is given by the set of derivatives $\omega_{\pm}=\{\partial_z,\partial_z^2,\cdots\}$ evaluated at a particular point, say $z=\frac{1}{2}$. Here however we are interested in another set, one which leads to optimal bounds on the OPE coefficients $a_{\Delta}$. They were constructed in  \cite{Paulos:2019gtx, Mazac:2016qev,Mazac:2018mdx, Mazac:2018ycv} and allow for the following basis decompositions:
	\be\label{basis}
	F_{\pm,\D} (z)=\sum_n \a_{\pm,n}^{B,F}(\D) F_{\pm,\D^{B,F}_n}(z)+\sum_n \b_{\pm,n}^{B,F}(\D)\partial F_{\pm,\D^{B,F}_n}(z)\,.
	\ee 
	The notation means here that we are free to either pick (B)osonic type functionals or (F)ermionic ones. The reason for this name has to do with the fact that the functionals satisfy certain duality conditions. For instance:
	\begin{align}
	\a_{\pm,n}^B(\D^B_m)&=\d_{mn},& \partial \a_{\pm,n}^B(\D^B_m)&=-c_{\pm,n}\d_{m0}\,,\nonumber\\
	\b_{\pm,n}^B(\D^B_m)&=0,& \partial \b_{\pm,n}^B(\D^B_m)&=\d_{mn}-d_{\pm,n}\d_{m0}\,.
	\end{align}
	Here $c_{\pm,n}$ and $d_{\pm,n}$ are some known coefficients and $\b_{\pm,n}^B=0$. For the fermionic functionals $\a_{\pm,n}^F$ and $\b_{\pm,n}^F$ we have a similar set of conditions with $\D_m^B\to \D_m^F, c_{\pm,n}\to 0,  d_{\pm,n}\to 0$, while all of them exist for all $n\ge 0$. 
	
	Completeness of the functional basis means that their action on the bootstrap equation gives a set of constraints which are necessary and sufficient for the validity of the bootstrap equation. In other words,
	\be
	\sum a_\D F_{\pm,\D}(z)=0 \Leftrightarrow \sum_\D a_\D \a_{\pm,n}(\D)=0\,, \ \ \sum_\D a_\D \b_{\pm,n}(\D)=0\,, \quad \forall n\geq 0.
	\ee
	It is understood in this equivalence that we may freely choose as complete sets either bosonic or fermionic functionals. We should mention that the '+'-functional basis,  constructed in \cite{Paulos:2019gtx}, has not actually been proven to be complete although it is likely this can easily following the same argument for the '-'-type basis \cite{Mazac:2018ycv}. This is because the fundamental ingredient for that proof is to have established that OPE coefficients are bounded parametrically by the free OPE density \reef{eq:opefree}, a fact which we will prove below.
	
	\subsection{Charged case}
	\label{ONsimp}
	
	In the presence of a global symmetry the crossing constraints are given in equation \ref{eq:crossingeq}. In this case we have not one but a set of $r$ crossing equations, each labelled by an index corresponding to an exchanged irrep. We define the action of a linear functional on a crossing vector $\mathcal F_{\Delta}^{\mf b}$ by:
	\bea
	\omega[\mathcal F^{\mf b}_{\Delta}]\equiv \omega(\mf b, \Delta):= \sum_{\mf a} \omega^{|\mf a}[\mathcal F_{\Delta}^{\mf b|\mf a}]\,.
	\eea
	The notation $\omega^{|\mf a}$ serves to emphasize that this is the $\mf a$-th component of the functional $\omega$, as we will shortly introduce functionals which are also labeled by a letter in this set.
	Furthermore we can write
	\bea
	\omega^{|\mf a}:=\sum_{\mf b} \left[P_{+}^{\mf b \mf a} \omega_-^{|\mf b}+P_{-}^{\mf b \mf a} \omega_+^{|\mf b}\right] \qquad \Rightarrow \qquad \omega(\mf b,\Delta)=\sum_{\mf a} \left[P^{\mf a \mf b}_+ \omega_-^{|\mf a}(\Delta)+P^{\mf a \mf b}_- \omega_+^{|\mf a}(\Delta)\right]
	\eea
	where $\omega_{\pm}^{|\mf a}$ are arbitrary functionals which act on $F_{\pm,\Delta}$. Each such functional leads to a sum rule:
	\bea
	\sum_{\mf b} a_{\Delta}^{\mf b} \omega(\mf b, \Delta)=0\,.
	\eea
	We will now use the bases of functionals described in the previous subsection to obtain a very simple basis of functionals for the global symmetry case. To do this we merely use expression \reef{eq:crossingeq} for the crossing vector $\mathcal F_{\Delta}^{\mf b}$ and plug in the bases decompositions for $F_{\pm,\Delta}$:
	\begin{multline}
	\mathcal F_{\Delta}^{\mf b|\mf a}=P^{\mf a \mf b}_{+}\sum_{n=0}^{+\infty}\left[\alpha_{-,n}(\Delta) F_{-,\Delta_n}+\beta_{-,n}(\Delta) \partial_{\Delta} F_{-,\Delta_n}\right]\\+P^{\mf a \mf b}_{-}\sum_{n=0}^{+\infty}\left[\alpha_{+,n}(\Delta) F_{+,\Delta_n}+\beta_{+,n}(\Delta) \partial_{\Delta} F_{+,\Delta_n}\right]\,.
	\end{multline}
	This can be rewritten as 
	\bea
	\mathcal F_{\Delta}^{\mf b|\mf a}=\sum_{\mf c}\sum_{n=0}^{\infty}\left[\widehat{\alpha}_n^{\mf c}(\mf b, \Delta) \mathcal F_{\Delta_n}^{\mf c|\mf a}+\widehat{\beta}_n^{\mf c}(\mf b, \Delta) \partial_{\Delta}\mathcal F_{\Delta_n}^{\mf c|\mf a}\right]\label{eq:expansionsimple}\,.
	\eea
	To see this we have defined
	\bea
	\widehat{\omega}_n^{\mf c}[\mathcal F^{\mf b}]=\sum_{\mf a} \widehat{\omega}_n^{\mf c|\mf a}[\mathcal F^{\mf b|\mf a}]\,, \qquad \widehat{\omega}_n^{\mf c|\mf a}=P^{\mf c \mf a}_+ \omega_{-,n}+P^{\mf c \mf a}_- \omega_{+,n}\,, \qquad \omega=\alpha,\beta
	\eea
	which gives in particular
	\bea
	\omega_n^{\mf c}(\mf b,\Delta)=P_+^{\mf c \mf b} \omega_{-,n}(\Delta)+P_-^{\mf c \mf b} \omega_{+,n}(\Delta)\,.
	\eea
	If we use $+,-$ functionals of the same kind, say both fermionic, this gives the duality relations
	\ba
	\widehat \alpha_n^{\mf a}(\mf b,\Delta_m^F)&=\delta_{n,m} \delta^{\mf a \mf b}\,,& \qquad \partial_{\Delta} \widehat \alpha_n^{\mf a}(\mf b,\Delta_m^F)&=0\\
	\widehat \beta_n^{\mf a}(\mf b,\Delta_m^F)&=0 \,,& \qquad \partial_{\Delta} \widehat \beta_n^{\mf a}(\mf b,\Delta_m^F)&=\delta_{n,m} \delta^{\mf a \mf b}\,,
	\ea
	These results strongly suggest that the set of functionals $\alpha_n^{\mf a}, \beta_n^{\mf b}$ (either bosonic of fermionic) form a complete basis of functionals which act on the crossing equation, with associated set of sum rules
	\bea
	\sum_{\mf b}\sum_{\Delta} a^{\mf b}\widehat \alpha_n^{\mf c}(\mf b,\Delta)=0\,, \quad \sum_{\mf b}\sum_{\Delta} a^{\mf b}\widehat \beta_n^{\mf c}(\mf b,\Delta)=0\,, \qquad \mbox{for all}\quad n\in \mathbb Z_{\geq 0}\,, \mf c\in \mf r \otimes \mf r\,. \label{eq:simplesumrules}
	\eea
	It is amusing to point out that these sum rules equivalently arise by demanding that a version of the Polyakov bootstrap holds,
	\bea
	\mathcal G^{\mf a}(z)=\sum_{\Delta} a^{\mf a}_{\Delta} G_{\Delta}(z)=\sum_{\mf b}\sum_{\Delta} a^\mf b_{\Delta} \widehat{\mathcal P}^{\mf b|\mf a}(z)
	\eea
	where the crossing symmetric ``simple'' Polyakov blocks have decompositions:
	\bea
	\widehat{\mathcal P}^{\mf b|\mf a}(z)=\delta^{\mf a \mf b} G_{\Delta}(z)-\sum_{n=0}^{\infty} \left[\widehat \alpha_n^{\mf a}(\mf b,\Delta) G_{\Delta_n}(z)+\widehat \beta_n^{\mf a}(\mf b,\Delta) \partial_{\Delta}G_{\Delta_n}(z)\right]\,.
	\eea
	where we can choose $\Delta_n=\Delta_n^{B,F}$ depending on the basis.
	This set of functionals is somewhat unusual, since it is dual to basis of operators which has the same set of scaling dimensions in each and every channel $\mf b$. This is certainly not consistent with our expectations for a basis of functionals associated to generalized free fields, where these dimensions do depend on the representation as mentioned in equation~\reef{eq:dimsgff}. Accordingly, the Polyakov blocks above are not computable, or at least not immediately, as a sum of AdS exchange Witten diagrams.
	
	A related way of understanding these functionals is as follows. In section \ref{sec:kinematics} we saw that any crossing-symmetric correlator with global symmetry can be recast as a decoupled set of $r_+$ crossing-symmetric and $r_-$ crossing-antisymmetric functions. Each such function satisfies a crossing equation, and we can introduce separate functional bases for each one of them. Indeed a generic functional can be written:
	\bea
	\omega^{|\mf a}=\sum_{s=1}^{r_+} e_{+,s}^{\mf a} \omega_{-}^s+\sum_{s=1}^{r_-} e_{-,s}^{\mf a} \omega_{+}^s
	\eea
	which applying to \reef{eq:crossingeq} gives
	\ba
	\sum_{\Delta} a_{+,s,\Delta} \omega_{-}^s(\Delta)&=0\,,\qquad s=1,\ldots, r_+\\
	\sum_{\Delta} a_{-,s,\Delta} \omega_{+}^s(\Delta)&=0\,,\qquad s=1,\ldots,r_-\\
	\ea
	i.e. the same as applying separate functionals to the crossing equations  \reef{eq:decoupledcrossing}.
	This makes it clear that we can use any functional basis we desire for each of these separate equations. In the simple functional basis above we have effectively made the, well, simplest possible choice, by using the same basis for all the $r_-$ equations, the same basis for all $r_+$ equations, and further setting the $r_-$ and $r_+$ basis to be both dual to fermionic (or bosonic fields). This is amounts to one possible choice of basis out of $2^{r}$ that could have been constructed in the same fashion.
	
	\subsection{OPE bounds}
	\label{sec:opebounds}
	Using the simple functional basis we can easily promote constraints on the uncharged OPE density to the charged ones. In particular consider a particular sum rule on the former taking the form:
	\bea
	\sum_{\Delta} a_{+,\Delta} \omega_-(\Delta)=0
	\eea
	with
	\bea
	a_{+,\Delta}:=\sum_{\mf a} e^{\mf a}_+ a^{\mf a}_{\Delta}\,, \qquad \sum_{\mf b} C^{\mf a \mf b} e_{+}^{\mf b}=e_{+}^{\mf b}
	\eea
	We will consider $\omega_-(\Delta)$ to be combinations of $\alpha_{-,n}, \beta_{-,n}$ which have nice positivity properties. To obtain interesting sum rules we have to ensure positivity of $a_{+,\Delta}$, but unitarity only gives $a^{\mf a}_\Delta\geq 0$. Hence we should choose crossing-matrix eigenvectors $e_+^{\mf b}$ with positive coefficients, preferrably as few as possible so that our bounds will be more constraining. We have defined one such eigenvector in equation \reef{eq:allposguy}, but we can do better. Indeed consider
	\bea
	E_{+}^{\mf a}:=\frac{d_{\mf r}}2\left(\delta^{\mf a \mbf S}+C^{\mf a \mbf S}+\eta^{\mf a} C^{\mf a \mbf S}\right)\,,\quad \tilde E_{+}^{\mf a}:=\frac {d_{\mf r}}2\left(\delta^{\mf a \mbf S}+C^{\mf a \mbf S}-\eta^{\mf a} C^{\mf a \mbf S}\right)\,.
	\eea
	Both of these are $+1$ eigenvectors with non-negative components, but furthermore the first is zero for $\eta^{\mf a}=-1$, while the second is zero for $\eta^{\mf a}=1$, with the exception $\mf a =\mbf S$. In detail,
	\bea
	E_+^{\mf a}=\left\{
	\begin{array}{lr}
		\frac{2+d_{\mf r}}2\,,& \mf a=\mbf S\\
		\sqrt{d_{\mf a}}\,, &	\eta^{\mf a}=+1\\
		0\,, &	\eta^{\mf a}=-1
	\end{array}
	\right.\,, \qquad
	\tilde E_+^{\mf a}=\left\{
	\begin{array}{lr}
		\frac{d_{\mf r}}2\,,& \mf a=\mbf S\\
		\sqrt{d_{\mf a}}\,, &	\eta^{\mf a}=-1\\
		0\,, &	\eta^{\mf a}=+1
	\end{array}
	\right.
	\eea
	Recall there are $r_+$ crossing symmetric eigenvectors overall, all of which can be chosen to have non-negative. components. By deforming $E_+^{\mf a}$ above, we see that among these we can always choose $r_+-r_-$ to have positive components and furthermore satisfy
	\bea
	\sum_{\mf b}(\delta^{\mf a \mf b}-\eta^{\mf a \mf b}) E_{+,s}^{\mf a}=0, \qquad s=1,\ldots, r_+-r_-
	\eea
	i.e. that they have all parity odd components equal to zero.
	The remaining $r_-$ vectors will necessary overlap with both parity odd and parity even irreps, although these overlaps can always be chosen positive.
	
\begin{figure}[t]
		\begin{center}
				\includegraphics[width=8cm]{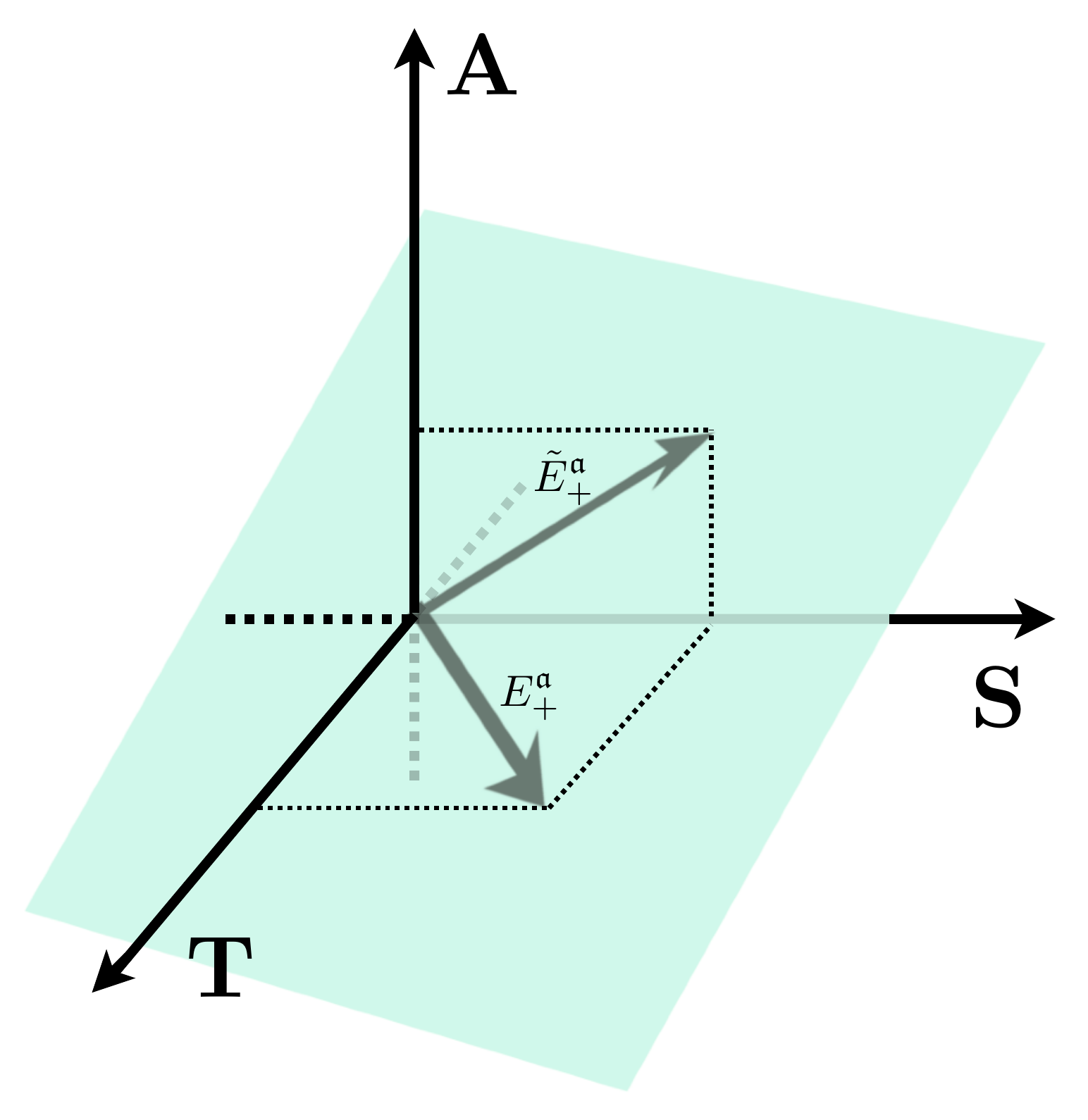}
			\caption{\label{fig:eigenvectors} Choice of $+1$ eigenvectors. For $O(N)$, where $r_+=2$ and $r_-=1$, the $r_+$ eigenvectors (green) of the crossing matrix with eigenvalue +1 can be chosen with all non-negative components. We can choose $1=r_+-r_-$ eigenvector with parity odd components set to zero and the other $r_-$ eigenvector(s) $\tilde{E}_+^{\mf a}$ to have zero non-singlet parity even components.}
		\end{center}
	\end{figure}

	To make the discussion less abstract, we can consider the case of $\phi_i$ in the fundamental of $O(N)$ described previously (see figure \ref{fig:eigenvectors}). In this case $r_+=2$, and we can choose as our basis the two vectors $E_{+}^{\mf a}, \tilde E_{+}^{\mf a}$ written above. In fact, these vectors have already appeared in section \ref{sec:kinematics}, where they are proportional to respectively  $E_{+,1}^{\mf a}$ and $E_{+,2}^{\mf a}$ written there.
	
	The conclusion is that suitable choices of vectors $E_+^{\mf a}$ allow us to translate any known bounds on unitary solutions to crossing without global symmetry to ones with it. Note that if a bound is saturated by some uncharged solution to crossing $\mathcal G_+(z)$, this can always be promoted to a solution to the full problem by simply setting
	\ba
	\mathcal G^{\mf a}(z)=E_+^{\mf a} \mathcal G_{+}(z)
	\ea
	This is clearly crossing symmetric, and since $\mathcal G_+(z)$ is unitary and $E_{+}^{\mf a}$ has positive components, so is the full $\mathcal G^{\mf a}(z)$. The only catch is that we should ensure that whatever solution we get should have the correct $\mbf S$ channel identity contribution.
	
	As a simple application, we can directly translate the OPE bounds found in reference \cite{Mazac:2018ycv}, where there was no global symmetry assumed, into the present context. By choosing $\omega_-$ to be suitable combinations of $\alpha_{-,n}$ and $\beta_{-,n}$ functionals, and with $e_+^{\mf a}$ an eigenvector positive components we obtain upper and lower bounds:
	
	\begin{align}
	\limsup\limits_{n\to \infty}\Bigg[\sum_{|\Delta_n-\D|\le 1}U_{\D,\D_n}\left(\frac{{\sum_{\mf b} e^{\mf b}_{+} a_{\D}^\mf b}}{a_{\D}^{\text{GFF}}}\right)\Bigg]&\le \sum_{\mf b}e^{\mf b}_{+,s} a_0^{\mf b}\,,\label{Uboundgen}\\
	\liminf\limits_{n\to \infty}\Bigg[\sum_{|\Delta_n-\D|\le 2}L_{\D,\D_n}\left(\frac{\sum_{\mf b} a_{\D}^{\mf b}e_{+}^{\mf b}}{a_{\D}^{\text{GFF}}}\right) \Bigg]&\ge \sum_{\mf b}e^{\mf b}_{+} a_0^{\mf b}\label{Lboundgen}\,.
	\end{align}
	with
	\be
	U_{\D,\D_n}=\frac{4 \sin^2\left[\frac{\pi }{2}(\D-2\D_n)\right]}{\pi ^2 (\Delta -\text{$\Delta_n$})^2}\,.
	\ee
	and 
	\be
	L_{\D,\D_n}=\frac{16 \sin^2\left(\frac{\pi }{2}(\D-2\D_n)\right) }{\pi ^2 (\Delta -\text{$\Delta_n$})^2(\D-\D_{n-1})(\D_{n+1}-\D)}
	\ee
	We are furthermore free to choose $\Delta_n$ to be $\Delta_n^B$ or $\Delta_n^F$. In particular, since we know there exists a choice of eigenvector with non-zero positive overlap with every irrep $\mf a$ (denoted $E_+^{\mf a}$ above), these imply that for any irrep the OPE density is bounded from above by the ``free'' one:
	\ba
	\limsup\limits_{n\to \infty}\Bigg[\sum_{|\Delta_n-\D|\le 1}U_{\D,\D_n}\left(\frac{{a_{\D}^\mf a}}{a_{\D}^{\text{GFF}}}\right)\Bigg]&\le \frac{\sum_{\mf b}E^{\mf b}_{+,i} a_0^{\mf b}}{E_{+}^{\mf a}}\,,\label{Uboundgen}
	\ea
	However, the same statement does not hold for the lower bound. Indeed, for any choice of irrep $\mf a$, other than the singlet, there are always unitary solutions to crossing where $\mathcal G^{\mf a}=0$: suffices to set $\mathcal G^{\mf a}$ proportional to $E^{\mf a}_{+}$ or $\tilde E^{\mf a}_{+}$ accordingly. 
	
	This means in particular that if we are interested in placing an upper bound on the dimension of the leading operator in any given irrep $\mf a$, this will always be infinite, except if $\mf a=\mbf S$. However, we can say that the joint gap in combinations of irreps is bounded. For instance, going back to our $O(N)$ example, the functionals:
	
	\ba
	\widehat \beta_{i}^{\mf a}:=E^{\mf a}_{+,i}\beta_{-,0}^F\,,
	\ea
	imply that setting a joint gap in the $\mbf S,\mbf T$ channel (while allowing identity contributions in both) is $1+2\Df$, and the same is true for the joint gap in the $\mbf S,\mbf A$ channels. In both cases the maximal gap is achieved by setting the full correlator proportional to the 1d generalized free fermion solution, e.g.
	\bea
	\mathcal G^{\mf a}(z)=E_{+,i}^{\mf a} \mathcal G^F(z)
	\eea
	where for $i=1$ the $\mbf A$ channel contribution is vanishing, while for $i=2$ the $\mbf T$ channel one is.

	\section{Interlude: Polyakov bootstrap}\label{sec:PolyakovBs}

		The goal of this section is to describe and explore the consequences of demanding an expansion for the correlator in terms of Polyakov blocks - crossing-symmetric functions associated to deformations of generalized free fields. Demanding this expansion holds will lead to a set of equations on the CFT data. These equations will be such that they will be trivially solved for generalized free fields. 
		
		We will show how Polyakov blocks can be defined in terms of AdS$_2$ Witten diagrams. In \cite{Ferrero:2019luz} (see also \cite{Dey:2016mcs} for earlier work) it was shown for the case of $O(N)$ that if the crossing symmetric basis just contains exchange Witten diagrams  the associated sum rules  are not convergent when OPE coefficients grow like that of a mean field theory. It was important then to add contact diagrams to get convergent sum rules. Further assuming that the OPE coefficients of unitary theories are parametrically bounded by the OPE coefficients of mean field theory, the exact structure of the crossing symmetric basis was found. 
		
		In this section we will systematically approach the problem for general global symmetries with similar arguments. In the next section we will show how to construct analytic functionals which lead to the same equations from a different, more rigorous, perspective, justifying the construction here.

	 	\subsection{Polyakov blocks and Witten diagrams}
	
	The Polyakov bootstrap states that the correlator may be expanded in a crossing-symmetric basis of functions, the Polyakov blocks. These come in bosonic and fermionic varieties. Here we will mostly restrict our discussion to the bosonic case since it's the one most familiar. We will comment on similarities and differences from the fermionic case as we go along.

	We postulate that a generic CFT correlator can be expanded as follows:
	\bea\label{Polboot}
	\mathcal G^{\mf a}(z)=\sum_{\Delta} a^{\mf a} G_{\Delta}(z|\Df)=\sum_{\mf b}\sum_{\Delta} a^{\mf b}_{\Delta} \mathcal P^{\mf b| \mf a}_\Delta(z)\,,
	\eea
	where the Polyakov blocks $\mathcal P^{\mf b}_{\Delta}$ are crossing symmetric,
	\bea
	\mathcal P^{\mf b|\mf a}_\Delta(z)=\sum_{\mf c} C^{\mf a \mf c} 	\mathcal P^{\mf b|\mf c}_\Delta(1-z)
	\eea
	and have the following conformal block decompositions:
	\bea
	\mathcal P^{\mf b| \mf a}_{\Delta}(z)=\delta^{\mf a \mf b} G_{\Delta}(z|\Df)-\sum_{n=0}^{\infty}\left[\alpha_n^{\mf a}(\mf b,\Delta) G_{\Delta_n^{\mf a}}(z|\Df)+\beta_n^{\mf a}(\mf b,\Delta_n^{\mf a}) \partial_{\Delta} G_{\Delta_n^{\mf a}}(z|\Df)\right]\,. \label{eq:polydecomposition}
	\eea
	As we will see in the next section, the coefficient functions $\alpha,\beta$ appearing above can be computed as functional actions.
	 Demanding \reef{Polboot} leads to the sum rules
	\begin{equation}
		\boxed{
			\sum_{\mf b}\sum_{\Delta} a^{\mf b}\alpha_n^{\mf c}(\mf b,\Delta)=0\,, \quad \sum_{\mf b}\sum_{\Delta} a^{\mf b} \beta_n^{\mf c}(\mf b,\Delta)=0\,, \qquad \mbox{for all}\quad n\in \mathbb Z_{\geq 0}\,, \mf c\in \mf r \otimes \mf r} \label{eq:gffsumrules}
	\end{equation}
	 This is similar to \reef{eq:simplesumrules}, but there is an important difference. Indeed, although these Polyakov blocks resemble the ones introduced in the previous section, notice that  in the expansion the scaling dimensions $\Delta_n^{\mf a}$ now depend on the channel $\mf a$. This suggests the functional actions now satisfy duality conditions:
	\ba\label{ortho}
	\alpha_n^{\mf a}(\mf b,\Delta_m^{\mf b})&\overset{?}{=}\delta_{n,m} \delta^{\mf a \mf b}\,,& \qquad \partial_{\Delta} \alpha_n^{\mf a}(\mf b,\Delta_m^{\mf b})&\overset{?}{=}0\\
	\beta_n^{\mf a}(\mf b,\Delta_m^{\mf b})&\overset{?}{=}0\,,& \qquad \partial_{\Delta} \beta_n^{\mf a}(\mf b,\Delta_m^{\mf b})&\overset{?}{=}\delta_{n,m} \delta^{\mf a \mf b}\,,
	\ea
	As it turns out these are almost correct but not quite. The correct expressions, and the reasons for them, will be discussed in the next subsection. 
	
\begin{figure}[t]
		\begin{center}
			\includegraphics[width=16cm]{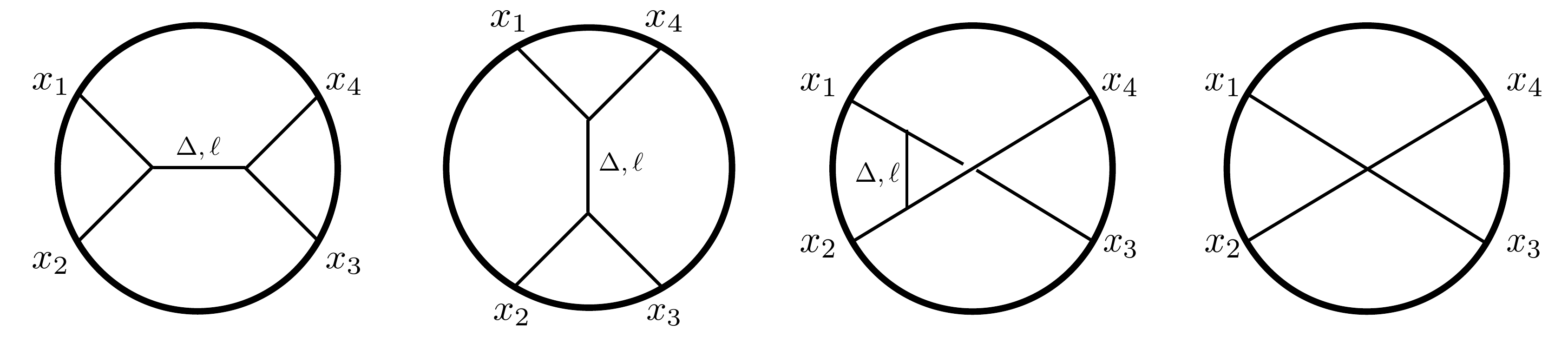}
			\caption{\label{fig:Wittendiagrams} Here we show the Witten exchange diagrams (exchanging a bulk operator of dimension $\Delta$ and spin $\ell=0,1$) in AdS$_2$ out of which a Polyakov block is built. The fourth diagram indicates specific (crossing symmetric) contact diagram contributions which need to be added to the Polyakov block for reasons discussed in section \ref{sec:subtract}. }
		\end{center}
	\end{figure}

	Let us now show that the Polyakov blocks actually exist, and obtain expressions for the functional actions appearing above. To see this we will need to consider Witten exchange diagrams in AdS$_2$, as illustrated in figure \ref{fig:Wittendiagrams}. These come in four different diagrams which can be thought of as four bosons/fermions exchanging a scalar/pseudoscalar state. Let us focus on the four boson case. The two possible Witten diagrams arise as dimensional reductions of the spin $\ell=0$ and spin $\ell=1$ exchanges in higher dimensions. They have decompositions
	\ba\label{decom1}
	W^{(s)}_{\Delta,\ell}(z)&=G_{\Delta}(z|\Df)+\sum_n \left[a_{n,\ell}^{(s)}(\D) G_{\D_{n,\ell}}(z|\Df)+ b_{n,\ell}^{(s)}(\D) \partial_{\Delta} G_{\D_{n,\ell}}(z|\Df)\right]\,, \qquad \ell=0,1
	\ea
	where $\Delta_{n,0}=2\Df+2n$ and $\Delta_{n,1}=2\Df+2n+1$. We can define their cross-channel versions by setting: 
	\bea\label{cross}
	W^{(t)}_{\Delta,\ell}(z)=W^{(s)}_{\Delta,\ell}(1-z),\hspace{0.5cm}W^{(u)}_{\Delta,\ell}(z)=(1-z)^{-2\D_\phi}\mbox{Re}\, W^{(s)}_{\Delta,\ell}\left(\frac{1}{1-z}\right)\,.
	\eea
 An important point is
	\bea
	W^{(u)}_{\Delta,\ell}(1-z)=(-1)^{\ell} W^{(u)}_{\Delta,\ell}(z)
	\eea
	In terms of these we define
	\be\label{plusminus}
	W^{(+)}_{\Delta,\ell}=\frac{W^{(t)}_{\Delta,\ell}+(-1)^{\ell} W^{(u)}_{\Delta,\ell}}{2}\,, \hspace{1cm} W^{(-)}_{\Delta,\ell}=\frac{W^{(t)}_{\Delta,\ell}-(-1)^{\ell} W^{(u)}_{\Delta,\ell}}{2}\,.
	\ee
	These definitions were chosen since they lead to simple block expansions:
	\ba\label{decom2}
	W^{(+)}_{\Delta,\ell}(z)&=\sum_{n}\left[ a_{n,\ell}^{(t)}(\Delta) G_{\D_n^B}(z)+ b_{n,\ell}^{(t)}(\Delta) \partial G_{\D_n^B}(z)\right], \\ 
	W^{(-)}_{\Delta,\ell}(z)&=\sum_{n}\left[ \bar a_{n,\ell}^{(t)}(\Delta) G_{\D_n^F}(z)+ \bar b_{n,\ell}^{(t)}(\Delta) \partial G_{\D_n^F}(z)\right]\,.
	\ea
	Let us now set:
	\bea
	W^{(s,t,u)}_{\Delta,\mf b}(z)=\left \{ \begin{array}{ll}
		W^{(s,t,u)}_{\Delta,0}(z) & \mbox{for}\ \eta^{\mf b}=1\\
		&\\
		W^{(s,t,u)}_{\Delta,1}(z) & \mbox{for}\ \eta^{\mf b}=-1
	\end{array}
	\right.
	\eea
	In terms of these we can finally write down our expression for the Polyakov block:
	\be\label{Poldef}
	\mathcal P^{\mf b}_{ijkl}(z)=W_{\D,\mf b}^{(s)}(z)\, T^{\mf b}_{ij,kl}+W_{\D,\mf b}^{(t)}(z)\, T^{\mf b}_{il,kj}+W_{\D,\mf b}^{(u)}(z)\, T^{\mf b}_{ik,jl}\,.
	\ee
	This can be alternatively written in components as follows:
	\bea
	\mathcal P^{\mf b|\mf a}(z)=\delta^{\mf a \mf b} W^{(s)}_{\Delta,\mf b}+C^{\mf a \mf b}\left[ W^{(t)}_{\Delta,\mf b}+\eta^a \eta^b W^{(u)}_{\Delta, \mf b}\right]\,.
	\eea
	It can be easily checked not only that this expression is crossing invariant but that it also has a decomposition of the form \reef{eq:polydecomposition}, with
	\bea
	\alpha_n^{\mf a}(\mf b,\Delta)=\left\{\begin{array}{ll}
		\delta^{\mf a \mf b} a_{n,0}^{(s)}(\Delta)+2\,C^{\mf a \mf b} a^{(t)}_{n,0}(\Delta) & \mbox{for}\ \eta^{\mf a}=+1\,, \eta^{\mf b}=+1\\
		\delta^{\mf a \mf b}a_{n,1}^{(s)}(\Delta)+2\,C^{\mf a \mf b} a^{(t)}_{n,1}(\Delta) & \mbox{for}\ \eta^{\mf a}=-1\,, \eta^{\mf b}=-1\\
		2\,C^{\mf a \mf b} \bar a^{(t)}_{n,0}(\Delta) & \mbox{for}\ \eta^{\mf a}=-1\,, \eta^{\mf b}=+1\\
		2\,C^{\mf a \mf b} \bar a^{(t)}_{n,1}(\Delta) & \mbox{for}\ \eta^{\mf a}=+1\,, \eta^{\mf b}=-1\\
	\end{array}
	\right.\label{eq:funcactionwitten}
	\eea
	and analogously for the $\beta$ functional actions. We discuss how the coefficients above can be computed in practice in appendix \ref{app:witten}.
	
		At this point we should make an important remark. The coefficients appearing in the conformal block decompositions above are actually not uniquely defined. One way to think about this ambiguity is that there exist solutions to crossing arising from AdS diagrams without any intermediate field being exchanged, also known as contact diagrams. Equivalently, the three point couplings between external states and the exchanged state labeled by $\Delta,\ell$ appearing in the Witten diagrams are also not uniquely defined, since one could always include derivatives in the three point coupling terms in the lagrangian. In the next section we will construct functionals $\alpha_n^{\mf a}, \beta_n^{\mf a}$ which will allow us to independently compute the quantities $\alpha_n^{\mf a}(\mf b,\Delta),\beta_n^{\mf a}(\mf b,\Delta)$ appearing in \reef{eq:funcactionwitten} as the action of those functionals on the crossing equation. Clearly the two different ways of computing these coefficients can only ever match if the ambiguities mentioned above are somehow fixed. Let us now describe how this is achieved.
		
	\subsection{Contact terms and subtractions}
		\label{sec:subtract}
	The ambiguities mentioned above can be expressed as crossing symmetric 4-point contact diagrams. 
	Considering the bosonic case for definiteness, the contact diagrams are Witten diagrams in AdS$_2$ built of 4 external $\Phi$ fields and a quartic vertex. The most general vertex is a linear combination of quartic interactions of the schematic form $\Phi_i\Phi_j(\partial^2)^m\Phi_k\Phi_l T^{\mf c}_{ij,kl}$. Such a contact diagram leads to a boundary correlator with a conformal block decomposition of the form
	\be
	\mathcal{C}^{|\mf a}(z)=\sum_n \big[a_{n,\mathcal{C}}^{\mf{a}} G_{\D_n^{\mf a}}(z)+ b_{n,\mathcal{C}}^{\mf{a}} \partial G_{\D_n^{\mf a}}(z)\big]\,,\label{eq:contactblocks}
	\ee 
	An important property of CFT correlators in unitary theories is boundedness in the $u$-channel Regge limit, where $z\to i \infty$.
	For the bosonic case the above contact diagrams are bounded in the Regge limit (i.e.\,$\sim z^\eta\,, \ \eta\le 0$), only for $m=0,1$ \cite{Gopakumar:2018xqi,Mazac:2018ycv}.\footnote{This can be seen from the Mellin amplitudes. In general if the Mellin amplitude behaves like $s^j$, in physical space the falloff should be $z^{j-1}$ \cite{Mazac:2018ycv}. In the limit of large Mellin variable $s$, the amplitudes corresponding to $m=0$ and $m=1$ vertices go like $s^{0}$ and $s^{1}$ respectively.}
	For fermionic external legs only contact diagrams built  with zero derivative vertices are Regge bounded \cite{Mazac:2018qmi,Mazac:2018ycv}. As we show in appendix \ref{app:contact}, it turns out that for bosonic external legs the exact number of contact diagrams that are Regge bounded is $r_+$, and for fermionic external legs it is $r_-$. For instance, for fermionic terms this is because each term is labeled by one of the $r_-$ parity odd tensor structures. 
	
	Consider now the Polyakov bootstrap equations \reef{eq:gffsumrules}. If the orthogonality conditions \eqref{ortho} were exactly correct, then it is easy to check these equations would imply that all the $a_{n,\mathcal C}, b_{n,\mathcal C}$ appearing in \reef{eq:contactblocks} would have to vanish. In particular this would exclude the 4-point Regge bounded contact diagrams. Since we want these solutions to be bootstrappable, it must be that some equations must be missing from the infinite set \eqref{eq:gffsumrules}. One convenient choice is to demand that the Polyakov blocks satisfy:
	\begin{align}\label{fixconds}
	\e=1: \ \ \ \ &\b_0^{\mf a}(\mf b,\D)=0 \ \text{ for } \  \eta^{\mf a}=1\,,\nonumber\\
	\e=-1: \ \ \ \ &\b_0^{\mf a}(\mf b,\D)=0 \  \text{ for } \ \eta^{\mf a}=-1\,.
	\end{align}
	This eliminates $r_+$ equations in the bosonic case and $r_-$ equations in the fermionic one, which matches the number of contact terms. 
	
	So how do we obtain Polyakov blocks satisfying these conditions? Let us denote a Witten diagram for a certain convenient choice of vertices as $\widetilde W^{(s,t,u)}_{\D,\mf b}(z)$, and the analogous uncorrected Polyakov block defined similar to \eqref{Poldef} as $\widetilde{\mathcal P}^{\mf b}(z)$.
	Then one relates the corrected Polyakov block to the uncorrected quantities as below:
	\be
	\mathcal{P}^{\mf b|\mf a}_{\Delta} = \widetilde{\mathcal{P}}^{\mf b|\mf a}_{\Delta}+\sum_{i=1}^{r_{\pm}} s_i^{\mf b}(\Delta) \mathcal{C}_{i}^{|\mf a} \  \text{ for } \ \e=\pm 1\,.
	\ee
	Here $\mathcal{C}_{i}$ are all the Regge bounded contact diagrams. The $r_{\pm}$ coefficients $s_i^{\mf b}(\Delta)$ above can then be chosen to impose \reef{fixconds}. More details on this procedure and the computation of the $\widetilde{W}^{(s,t,u)}_{\D,\mf b}(z)$ and associated block decomposition coefficients are given in appendix \ref{app:witten}.
	
	To summarize, we claim at the end of this procedure the sum rules \reef{eq:gffsumrules} hold, but some of the functional actions appearing there are identically zero. The duality conditions become:
	\ba
	\alpha_n^{\mf a}(\mf b,\Delta_m^{\mf b})&=\delta_{n,m} \delta^{\mf a \mf b}\,,& \qquad \partial_{\Delta}  \alpha_n^{\mf a}(\mf b,\Delta_m^{\mf b})&=-d_n^{\mf a, \mf b}\delta_{m,0}\\
	\beta_n^{\mf a}(\mf b,\Delta_m^{\mf b})&=0\,,& \qquad \partial_{\Delta}  \beta_n^{\mf a}(\mf b,\Delta_m^{\mf b})&=\delta_{n,m} \delta^{\mf a \mf b}-c_n^{\mf a, \mf b}\delta_{m,0} \label{eq:dualitygff1}
	\ea
	for some coefficients $c_n, d_n$ related to the conformal block decompositions of contact diagrams and the $s_i^{\mf b}$ above. Finally, the functional actions may be computed in terms of sums of Witten diagrams.

	\section{The GFF functional bases}\label{sec:GFFfunctionals}
	
	We will now show how to construct functional bases that bootstrap the Polyakov blocks presented in the previous section. The functional actions lead to the very same sum rules derived there, thereby justifying them more rigorously. Also, we will see that these bases are neatly encapsulated in master functionals \cite{Paulos:2020zxx}. The master functionals in turn lead to crossing symmetric dispersion relations for charged CFT correlators.
		
	\subsection{Fundamental free equation}
	We would like to define a general class of functionals suitable for acting on the crossing equation \reef{eq:crossingeq}. We will work with a simple generalization of the functionals in \cite{Mazac:2018mdx,Mazac:2018ycv}. We begin by setting
	\bea
	\omega(\mf b,\Delta)=\int_1^{\infty} \ud z \sum_{\mf a} h^{|\mf a}(z)  \mathcal I_z\mathcal F^{\mf b|a}(z)\,,
	\eea
	with $\mathcal I_z \mathcal F(z)=\lim_{\epsilon\to 0} \mbox{Im} \mathcal F(z+i\epsilon)$. By deforming the contour, we work with the more convenient definition:
	\bea
	\omega(\mf b,\Delta)=\frac 12 \int_{\frac 12}^{\frac 12+i\infty}\!\! \ud z\, \sum_{\mf a} f^{|\mf a}(z) \mathcal F^{\mf b|\mf a}(z) + \int_{\frac 12}^1 \ud z \sum_{\mf a} g^{|\mf a}(z) \mathcal F^{\mf b|\mf a}(z)
	\eea
	where
	\bea
	f^{|\mf a}(z)&=&\frac{h^{\mf a}(z)-\sum_{\mf b} C^{\mf b \mf a} h^{|\mf b}(1-z)}{i \pi}\,, \qquad \mbox{Im} z>0\\
	g^{|\mf a}(z)&=&-\frac{\mathcal I_z h^{|\mf a}(z)}{\pi}\,, \qquad z\in(0,1)
	\eea
	Note that we have
	\bea
	f^{|\mf a}(z)=\sum_{\mf b} C^{\mf b \mf a} f^{|\mf b}(1-z)\,.
	\eea
	These definitions imply the gluing condition:
	\bea
	\mathcal R_z f^{|\mf a}(z)=-g^{|\mf a}(z)-\sum_{\mf b} C^{\mf b \mf a} g^{|\mf b}(1-z)\,, \qquad z\in(0,1)\,.
	\eea
	We wish to constrain the kernels such that they will describe functionals satisfying duality conditions of the form \reef{eq:dualitygff1}. Let us take $f^{\mf a}(z)$ real for $z>1$ and $z<0$. Using the definition of $\mathcal F^{\mf b}$, a simple contour manipulation (whose validity we will comment on later), together with the above condition, leads to
	\begin{multline}
	\omega(\Delta,\mf b)=\int_0^1 \ud z\, g^{|\mf b}(z) G_{\Delta}(z|\Df)-\int_{-\infty}^0 \ud z f^{|\mf b}(z) \mathcal R_z G_{\Delta}(z|\Df)\\
	=\int_0^1\ud z\,\left[ g^{|\mf b}(z)-\cos[\pi(\Delta-2\Df)] (1-z)^{2\Df-2}\,f^{|\mf b}(\mbox{$\frac z{z-1}$})\right]\, G_{\Delta}(z|\Df)
	\end{multline}
	It follows that if we set
	\bea
	g^{|\mf b}(z)=\epsilon\eta^{\mf b}\, (1-z)^{2\Df-2}\,f^{|\mf b}(\mbox{$\frac z{z-1}$}) 
	\eea
	with $\epsilon=+1 (-1)$ for a boson (fermion), the functional action will have double zeros for $\Delta=\Delta^{\mf b}_n$:
	\bea
	\omega(\mf b,\Delta)=2\sin^2\left[\frac{\pi}2(\Delta-\Delta_0^{\mf b})\right]\int_0^1\, g^{|\mf b}(z)\, G_{\Delta}(z|\Df) \label{eq:funcactionpos}
	\eea
	where $\Delta_n^{\mf b}$ was defined in \reef{eq:dimsgff}. Plugging in the relation between $f^{|\mf b}, g^{|\mf b}$ into the gluing condition we find the fundamental equation:
	\begin{equation}
	\label{eq:fundfreeeq}
	\boxed{
		\epsilon\, \eta^{\mf a} \mathcal R_z f^{|\mf a}(z)=-(1-z)^{2\Df-2} f^{|\mf a}(\mbox{$\frac{z}{z-1}$})-z^{2\Df -2}\sum_{\mf b} \eta^{\mf a} C^{\mf b \mf a} \eta^{\mf b} f^{|\mf b}(\mbox{$\frac{z-1}{z}$})
	}
	\end{equation}
	Note that if we remove all reference to $\eta^{\mf a}$ in the above equation it becomes diagonal in the $\mf a$ index and in fact the equation then describes the simple functionals studied in section \ref{sec:simplefuncs}.
	
	\subsection{Boundary conditions and solutions}\label{sec:conditions}
	
	The solutions can be labeled by their behaviour near $z=0$. We can split the set of solutions into classes of functionals associated to individual representations, so that we specialize $f^{|\mf a} \to f^{\mf b|\mf a}$. First we define prefunctionals which satisfy  boundary conditions:
	\ba\label{zto0}
	\tilde \alpha_n^{\mf a|\mf b}:&\qquad& f^{\mf a|\mf b}(z)&\underset{z\to 0^-}\sim -\frac{2}{\pi^2}\, \frac{\eta^{\mf a}\delta^{\mf a \mf b}(\log(z)}{z^{1+\Delta^{\mf a}_n-2\Df}}\\
	\tilde \beta_n^{\mf a|\mf b}:&\qquad& f^{\mf a|\mf b}(z)&\underset{z\to 0^-}\sim \frac{2}{\pi^2}\, \frac{\eta^{\mf a}\delta^{\mf a \mf b}}{z^{1+\Delta^{\mf a}_n-2\Df}}
	\ea
	Note that these boundary condition still do not specify the kernels uniquely. For instance, in the above a solution with a given $n$ can always be shifted by solutions with lower $n$. This ambiguity can be fixed by demanding the duality conditions:
	\ba
	\tilde \alpha_n^{\mf a}(\mf b,\Delta_m^{\mf b})&=\delta_{n,m} \delta^{\mf a \mf b}\,,& \qquad \partial_{\Delta} \tilde \alpha_n^{\mf a}(\mf b,\Delta_m^{\mf b})&=0\\
	\tilde \beta_n^{\mf a}(\mf b,\Delta_m^{\mf b})&=0\,,& \qquad \partial_{\Delta} \tilde \beta_n^{\mf a}(\mf b,\Delta_m^{\mf b})&=\delta_{n,m} \delta^{\mf a \mf b}\,.
	\ea
	Concretely, the boundary conditions that we set at $z=0$ above automatically guarantees the duality conditions hold whenever $m\geq n$. This easily seen to follow from the representation \reef{eq:funcactionpos}. Imposing they also hold for $m<n$ is precisely what fixes the remaining ambiguity allowing us to shift functionals with a given $n$ by lower ones.
	
	However, there is an important catch, which is that the kernels corresponding to the prefunctionals $\tilde \alpha, \tilde \beta$ will in general not satisfy the correct fall-off at infinity. Indeed, the kernels $f^{|\mf a}$ should satisfy:
	\bea\label{ztoinf}
	f^{|\mf b}(z)\underset{z\to \infty}= O(z^{-2})
	\eea
	This requirement is unchanged relative to the case without global symmetry, and is necessary to ensure that the functionals are crossing compatible, i.e. that their actions commutes over the infinite sum over states in the crossing equations \cite{Mazac:2018mdx,Mazac:2018ycv}.
	
	The solution to this problem is to define appropriate finite linear combinations of prefunctionals. This behaviour is already familiar from the construction of the bosonic functional basis in the absence of global symmetries, and reflects the existence of simple solutions to crossing related to contact terms in AdS. What's new for global symmetries is that even for fermionic basis there are subtractions required. In all cases we've checked, these subtractions satisfy:
	\bea
	\left(\mbox{\# of subtractions}\right) \qquad = \qquad \left(\mbox{\# of irreps $\mf a$ with $\epsilon \eta^\mf a=1$}\right)
	\eea
	Notice this matches the expectations of section \ref{sec:subtract}.	We choose to always subtract $\tilde \beta$ functionals, such that:
	\ba\label{betasub}
	\beta_n^{\mf a}&= \tilde \beta_n^{\mf a}-\sum_{\mf c} c_n^{\mf a,\mf c} \tilde \beta_0^{\mf c}\\
	\alpha_n^{\mf a}&=\tilde \alpha_n^{\mf a}- \sum_{\mf c} d_n^{\mf a,\mf c} \tilde \beta_0^{\mf c}
	\ea
	where it is understood that both $c_n^{\mf a, \mf c}$ and $d_n^{\mf a, \mf c}$ are zero unless $\epsilon\eta^{\mf c} =1$. The subtractions modify the duality conditions:
	\ba
	\alpha_n^{\mf a}(\mf b,\Delta_m^{\mf b})&=\delta_{n,m} \delta^{\mf a \mf b}\,,& \qquad \partial_{\Delta}  \alpha_n^{\mf a}(\mf b,\Delta_m^{\mf b})&=-d_n^{\mf a, \mf b}\delta_{m,0}\\
	\beta_n^{\mf a}(\mf b,\Delta_m^{\mf b})&=0\,,& \qquad \partial_{\Delta}  \beta_n^{\mf a}(\mf b,\Delta_m^{\mf b})&=\delta_{n,m} \delta^{\mf a \mf b}-c_n^{\mf a, \mf b}\delta_{m,0}\,.\label{eq:correctduality}
	\ea
	Again this is in perfect agreement with the discussion of the previous section.
	
\begin{figure}[t]
		\begin{center}
			\begin{tabular}{lr}
			\includegraphics[width=7.5cm]{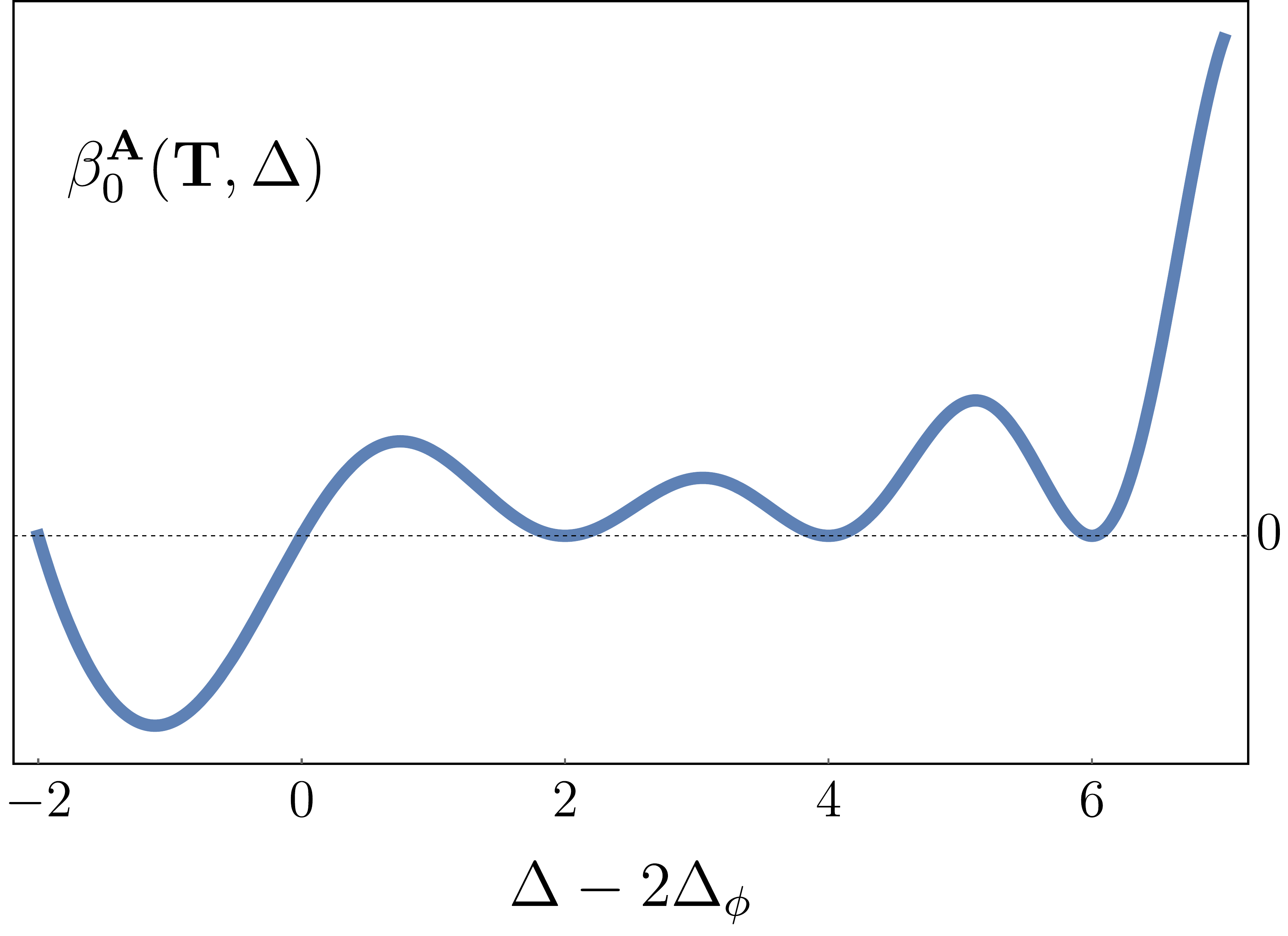}
			&
			\includegraphics[width=7.5cm]{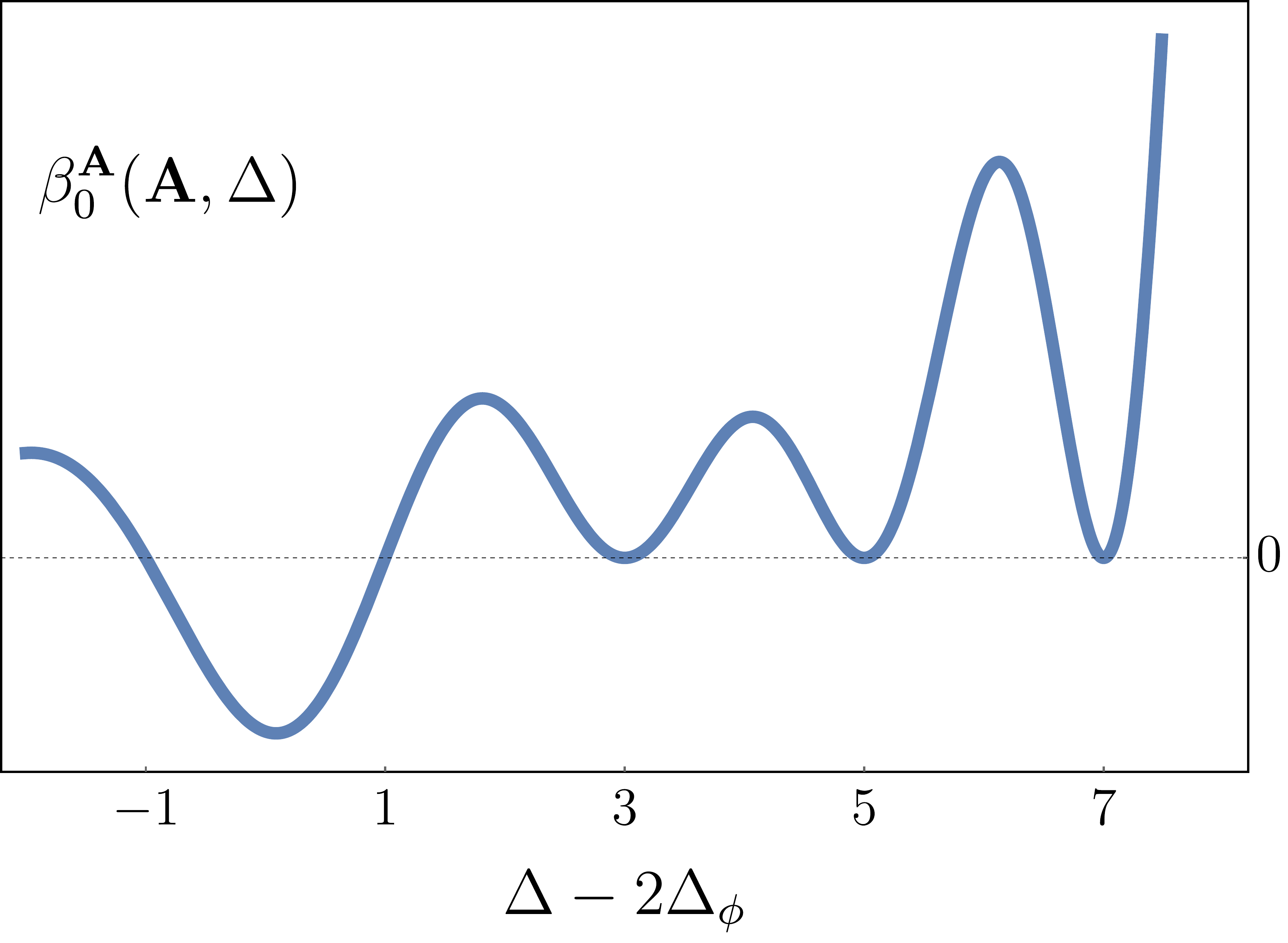}
			\end{tabular}
			\caption{\label{fig:b0} Bosonic functional $\beta^{\mbf A}_0$. We consider the $O(n)$ case with $n=3$ and plot its action in the $\mbf T$ and $\mbf A$ channels for $\Df=1$. The action in the $\mbf S$ channel is proportional to that in the $\mbf T$ channel.}
		\end{center}
	\end{figure}

	We have yet to present actual functional kernels satisfying all the desired boundary conditions. For special values of $\Df$, the construction of the functional kernels can be done very explicitly for the entire basis. For more general values, we can present an alternative basis of functionals which is equivalent to the one above after orthonormalization. Details are given in appendix \ref{app:gffkernels}. Two typical functionals are shown in figure \ref{fig:b0}.
		
	\subsection{Master functionals and dispersion relation}
	
	A convenient way to package the functional bases is by introducing master functionals \cite{Paulos:2020zxx}. In the present context we will define these as:
	\bea
	\Omega^{\mf a}_w:=\epsilon \eta^{\mf a} \sum_{n=0}^\infty \left[ G_{\Delta_n^{\mf a}}(w|\Df)\,\alpha_n^{\mf a}+\partial_{\Delta} G_{\Delta_n^{\mf a}}(w|\Df)\, \beta_n^{\mf a}\right]
	\eea
	where as usual there are really two kinds, depending on whether we consider the bosonic or fermionic basis. Their functional actions are given by
	\bea
	\epsilon \eta^{\mf a}\,\Omega^{\mf a}_w(\mf b,\Delta)=\delta^{\mf a,\mf b}\,G_{\Delta}(w|\Df)-\mathcal P_{\Delta}^{\mf b|\mf a}(w)
	\eea
	with associated sum rule
	\bea
	\sum_{\mf b} a^{\mf b}(\Delta) \Omega^{\mf a}_w(\mf b,\Delta)=0 \Leftrightarrow \mathcal G^{\mf a}(w)=\sum_{\mf b} \sum_{\Delta} a^{\mf b}_{\Delta} \mathcal P^{\mf b|\mf a}_\Delta(w).
	\eea
	That is, the master functional sum rule is equivalent to the validity of the Polyakov bootstrap. Since the master functional is, well, a functional, it can also be defined independently in terms of its own $f, g$ kernels. Setting
	\bea
	g_w^{\mf a|\mf b}(z)=\epsilon \eta^{\mf a}\delta^{\mf a,\mf b} \delta(w-z)+ \hat g_w^{\mf a|\mf b}(z)\,,\qquad \hat g_w^{\mf a|\mf b}(z)=\epsilon \eta^{\mf b} f_w^{\mf a|\mf b}(\mbox{$\frac{z}{z-1}$})\,,
	\eea
	the gluing condition of the previous section becomes:
	\begin{multline}
	\label{eq:mastereq}
	\epsilon\, \eta^{\mf b} \mathcal R_z f_w^{\mf a|\mf b}(z)+(1-z)^{2\Df-2} f^{\mf a|\mf b}(\mbox{$\frac{z}{z-1}$})+z^{2\Df -2}\sum_{\mf c} \eta^{\mf b} C^{\mf c \mf b} \eta^{\mf c} f^{\mf a|\mf c}(\mbox{$\frac{z-1}{z}$})\\=-\left[\delta^{\mf a,\mf b}\delta(w-z)+C^{\mf a \mf b} \delta(1-w-z)\right]
	\end{multline}
	In appendix \ref{app:master} we explain how this equation can be solved in practice. In cases where we can find an explicit analytic solution, we can reexpand the resulting kernels in $w$ to recover the $\alpha_n, \beta_n$ functionals finding perfect agreement.
	
	Using formula \reef{eq:funcactionpos} as applied to the master functionals, the corresponding sum rules can be recast as a dispersion relation for the correlator:
	\begin{equation}
	\sum_{\mf b} a^{\mf b}(\Delta) \Omega^{\mf a}_w(\mf b,\Delta)=0\Leftrightarrow
	\boxed{ \overline{\mathcal G}^{\mf a}(w)=-\epsilon \eta^{\mf a}\int_0^1 \ud z\, \sum_{\mf b} g^{\mf a|\mf b}_w(z) \,d^2 \overline{\mathcal G}^{\mf b}(z)
	}\,,
	\end{equation}
	In the above we have defined the double discontinuity: 
	\bea
	d^2\mathcal G^{\mf a}(z):=\mathcal G^{\mf a}(z)-\epsilon\eta^{\mf a}(1-z)^{-2\Df} \mathcal R_z \mathcal G^{\mf a}(z)(\mbox{$\frac z{z-1}$})\,.
	\eea
	The dispersion relation applies only to a suitably subtracted correlator, since representation \reef{eq:funcactionpos} is only valid for sufficiently large $\Delta$
	\bea
	\overline{\mathcal G}^{\mf a}(z):=\mathcal G^{\mf a}(z)-\sum_{\mf b_-}\sum_{\Delta\leq 2\Df-1} a^{\mf b}_{\Delta} \mathcal P^{\mf b|\mf a}_\Delta(z)-\sum_{\mf b_+}\sum_{\Delta\leq 2\Df} a^{\mf b}_{\Delta} \mathcal P_\Delta^{\mf b|\mf a}(z)
	\eea
	where $\mf b_{\pm}$ are representations for which $\epsilon \eta^{\mf b_\pm}=\pm 1$. The range of the subtractions is fixed by the boundary conditions satisfied by the master functional kernels. 
	
\begin{figure}[t]
		\begin{center}
			\begin{tabular}{lr}
				\includegraphics[width=7.5cm]{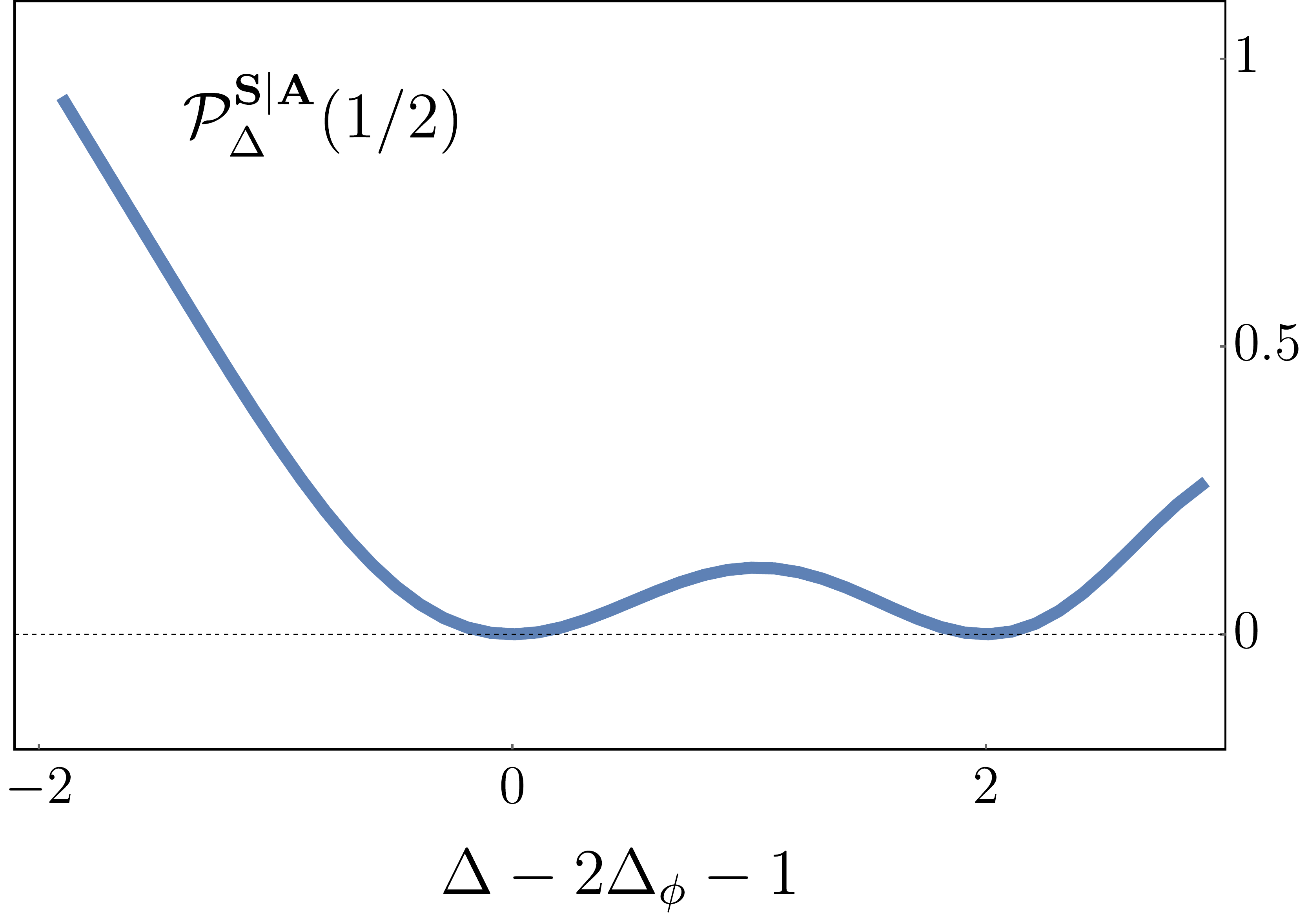}
				&
				\includegraphics[width=7.5cm]{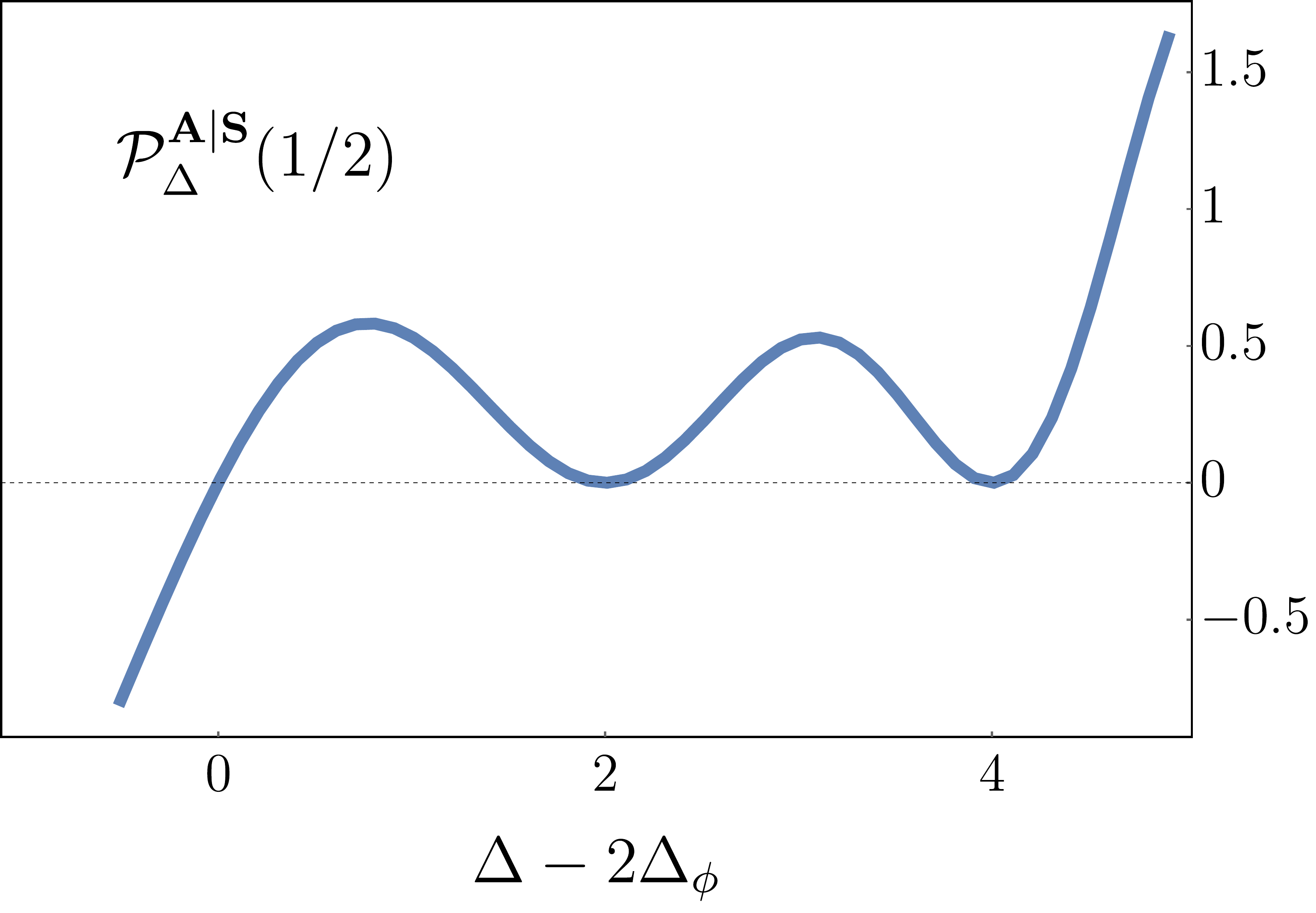}
			\end{tabular}
			\caption{\label{fig:mf} Fermionic master functional. We consider the $O(n)$ case with $n=3$ and plot representative components of the fermionic Polyakov blocks $\mathcal P_{\Delta}^{\mbf S}$ and $\mathcal P_{\Delta}^{\mbf A}$ for $\Df=\frac 12$ evaluated at the crossing symmetric point $z=1/2$.}
		\end{center}
	\end{figure}

	The dispersion relation is nothing but a restatement of the Polyakov bootstrap, since the Polyakov blocks themselves satisfy the dispersion relation. Indeed, using the property:
	\bea
	d^2 \mathcal P^{\mf b|\mf a}_\Delta(z)=2 \delta^{\mf a \mf b} \sin^2\left[\frac{\pi}2(\Delta-\Delta^{\mf b}_0)\right]\,G_{\Delta}(z|\Df)
	\eea
	gives
	\bea
	\mathcal P_{\Delta}^{\mf b|\mf a}(w)=2\epsilon \eta^{\mf a}\, \sin^2\left[\frac{\pi}2(\Delta-\Delta^{\mf b}_0)\right] \int_0^1 \ud z\, g^{\mf a|\mf b}_w(z)\,G_{\Delta}(z|\Df).
	\eea
	This formula can be used to compute the Polyakov blocks directly, as an alternative to their description in terms of Witten exchange diagrams. It also directly translates positivity properties of the master functional kernels into those of the Polyakov blocks. We show a sample computation of the blocks in figure \ref{fig:mf}.
	
	\section{Numerical applications}\label{sec:numerics}
	
	The GFF functionals can be used as a basis for numerical bootstrap applications. In this section we discuss two of these, to constrain correlators of a field transforming in the fundamental irrep of an $O(N)$ global symmetry. Firstly we obtain minimal and maximal bounds on the values of such correlators, and secondly, we revisit the bootstrap of the 3d Ising twist defect first considered in \cite{ Gaiotto:2013nva}.

	\subsection{The space of $O(N)$ correlators}

	We are interested in determining the range of allowed values for CFT correlators of fields transforming in the fundamental representation of $O(N)$ symmetry\footnote{A similar problem has been adressed for S-matrices in \cite{Cordova:2019lot}.}. 	 Recall that in this case there are three exchanged irreps, namely $\mbf S,\mbf T$ and $\mbf A$. The goal of this section is provide an initial exploration of this space, leaving a more detailed investigation for future work.
	
	There are at least two ways of formulating the problem. The first is to ask for the range of variation of $\mathcal G^{\mf a}$ when it lies along some chosen direction $n^{\mf a}$. We call this the radial formulation of the problem:
	\bea
    \mbox{Radial:}\quad \mbox{min (max)}\quad	R:=\sqrt{\sum_{\mf a}[\mathcal G^{\mf a}(z)]^2}\qquad \mbox{such that}\quad \mathcal G^{\mf a}(z)=R n^{\mf a}\,.
    \eea
	The second way is to again choose some vector $\theta^{\mf a}$ as the normal direction to some plane and ask what is the correlator which minimizes/maximizes the distance to that plane. We call this the plane formulation:
	\bea
	\mbox{Plane:}\quad \mbox{min (max)} \quad \sum_{\mf a}\theta^{\mf a}\mathcal G^{\mf a}(z)\,.
	\eea
	Note that unitarity means we should take $n^{\mf a}$ with all positive components. As for $\theta^a$, we must set its components to be all positive/negative when doing minimization/maximization. Either way, by varying over  $\vec n$ or $\vec \theta$ we expect to recover the same space of allowed correlator values. This space is bounded by some two dimensional surface embedded in $\mathbb R^3$, and whereas the plane method will give values clustered around points of high curvature on this surface, the first method is more democratic but might miss such points. In the following we will use a combination of the two. 
	
	Before we proceed to explain how the above may be formulated as linear optimization problems, let us discuss some basic expectations. Firstly, the analysis of section 
	\ref{sec:simplefuncs} tells us that a trivial set of solutions to the charged crossing equation can be obtained from the uncharged case by setting $\mathcal G^{\mf a}\propto E^{\mf a} \mathcal G(z)$, with $E^{\mf a}$ an eigenvector of the crossing matric $C^{\mf a\mf b}$ with non-negative components. For $O(N)$ there are two such vectors, which we repeat here for convenience:
	\bea
		E_{+,1}^{\mf a}=\left(1,\frac{N-1}{\sqrt{d_{\mbf T}}},0\right)\,, \quad  E_{+,2}^{\mf a}=\left(1,0,\frac{N-1}{\sqrt{d_{\mbf A}}}\right)\,.
	\eea
	This immediately tells us that the minimal allowed values for both $\mathcal G^{\mbf T}$ and $\mathcal G^{\mbf A}$ are actually zero. Furthermore, it also shows us that correlator values may become arbitrarily large unless we impose a gap in the spectrum. This is because for sufficiently small gap there are unitary solutions to the uncharged crossing equation without identity which may be included with an arbitrarily large coefficient. To deal with this, in the following we will impose a uniform gap in all three channels equal to $2\Df$ when addessing the maximization problem. However, we choose to still allow for identity contributions in individual channels. For convenience we set
	\ba
	\mbox{minimization:}&\quad& S&:=\{\Delta: \Delta\in [0,\infty)\}\\
	\mbox{maximisation:}&\quad& S&:=\{\Delta: \Delta\in \{0\}\cup [2\Df,\infty)\}		
	\ea
	In reference \cite{Paulos:2020zxx}, one of us established exact correlator bounds for the uncharged case. Following the analysis of section \ref{sec:simplefuncs} these translate in the current context to the bounds:
	\bea
\left(\sum_{\mf a} E_{+,i}^{\mf a}a_{0}^{\mf a}\right)\, \mathcal G^{F}(z)\leq	\sum_{\mf a} E_{+,i}^{\mf a} \mathcal G^{\mf a}(z) \leq \left(\sum_{\mf a} E_{+,i}^{\mf a}a_{0}^{\mf a}\right)\, \mathcal G^{B}(z)\,, \quad i=1,2\,.
	\eea
	Here $\mathcal G^F$ ($\mathcal G^B$) are given by%
	\bea
	\mathcal G^{B,F}(z)=\pm 1+z^{-2\Df}+(1-z)^{-2\Df}
	\eea
	and describe generalized free Boson (+1) or Fermion (-1) solutions. The bounds corresponding to $i=1$ will be saturated by choosing $\mathcal G^{\mf a}(z)=E_{+,2}^{\mf a} \mathcal G^{B,F}(z)$ and vice-versa.

	Hence, for these choices of $\theta^{\mf a}$ or $n^\mf a$ we already know what we will get.
	
	Let us now explain how to obtain bounds by constructing suitable functionals for each method. We first introduce the functional search space:
	\ba
	\Lambda_{n_{\mbox{\tiny max}}}:=\mbox{span}\left\{ \alpha_n^{\mf a},\beta_n^{\mf a}\,, \quad n=0,\ldots,n_{\mbox{\tiny max}}\,; \mf a=\mbf S, \mbf T,\mbf A\right\}
	\ea

\begin{figure}
		\centering
		\begin{tabular}{lr}
				\includegraphics[width=6.5cm]{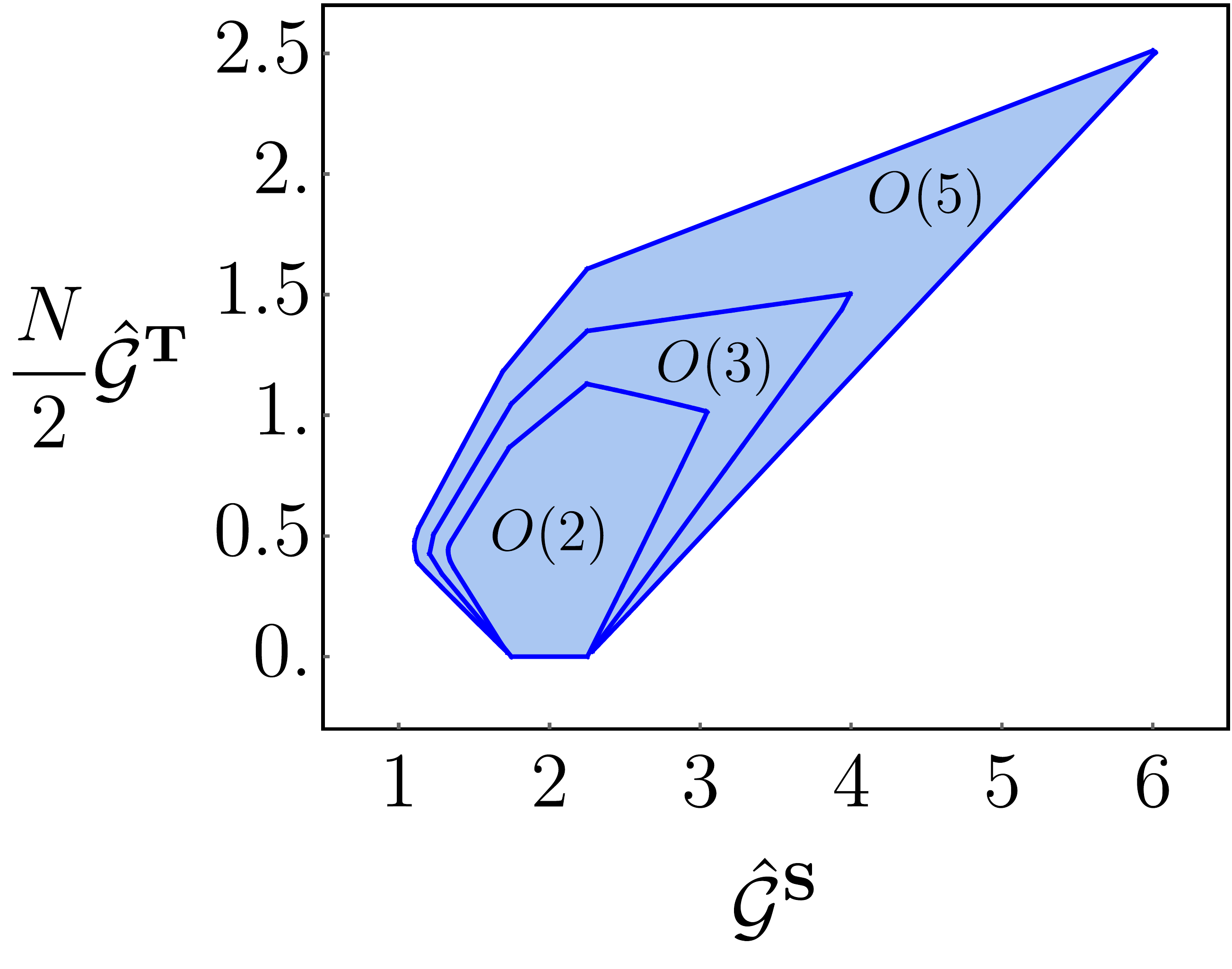}	
		&
		\includegraphics[width=8cm]{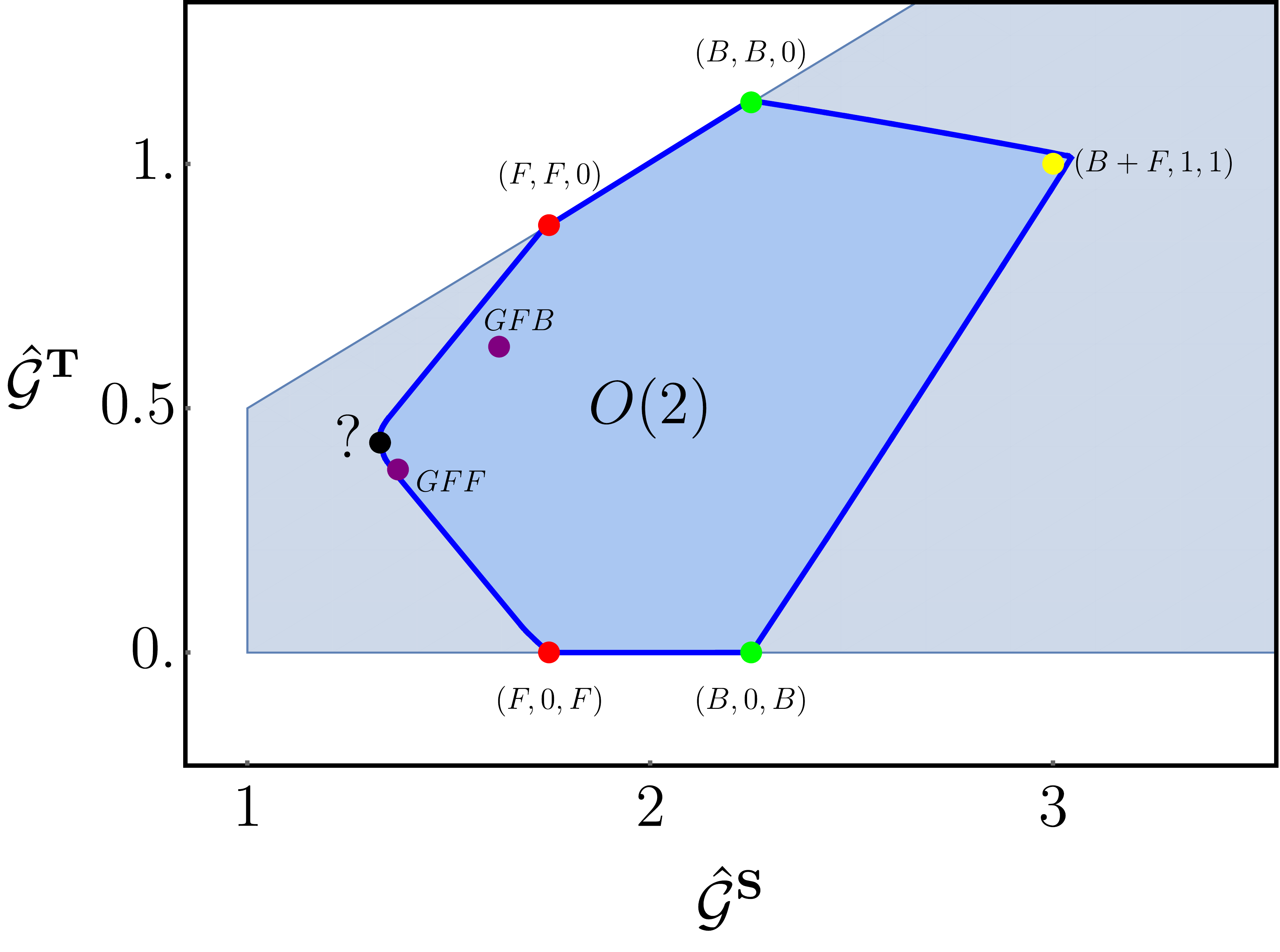}	
		\end{tabular}
		\caption{\label{fig:2d}Bounds on $O(N)$ correlators at the crossing symmetric point $z=1/2$ and $\Df=1$. We present results in terms of $\hat {\mathcal G}^{\mf a}(z):=z^{2\Df} \mathcal G^{\mf a}(z)/\sqrt{d_{\mf a}}$. On the left bounds for various values of $N$, on the right for $N=2$. The light shaded regions represent the trivial constraints $\hat{\mathcal G}^{\mf a}\geq \delta^{\mf a \mbf S}$. The colored marked points on the right represent special solutions to crossing, with GFF and GFB (violet) the O(2) generalized free fermion/boson. Other points are explained in the main text.
	}
	\end{figure}

	In this work we have used the bosonic GFF functionals with $n_{\mbox{\tiny max}}$ ranging from 2 to 6, i.e. we use bases containing up to 40 functionals\footnote{As explained in \cite{Paulos:2020zxx}, for the maximization problem we must include one additional functional, which here we choose to be an element of the fermionic GFF basis.}.
	In the plane method we solve the problem:
	\ba
	\underset{\Omega\in \Lambda_{ n_{\mbox{\tiny max}}}}{\mbox{max}}\ \Omega(\mbf S,0): \qquad \Omega(\mf a,\Delta)&\geq -\theta^{\mf a} G_{\Delta}(z|\Df)\,, \quad \Delta\in S
	\ea
	Depending on the signs of the components of $\theta^{\mf a}$ this captures both minimization as well as maximization problems. For instance, in the minimisation case where all $\theta^{\mf a}$ are non-negative, acting with such a functional on the crossing equation gives
	\ba
	\sum_{\mf a}\sum_{\Delta \in S}a_0^{\mf a}\Omega(\mf a,\Delta)=0 \Rightarrow \Omega(\mbf S,0)+\theta^{\mbf S}\,G_0(z) \leq \sum_{\mf a} \theta^{\mf a} \mathcal G^{\mf a}(z),.
	\ea
	
	For the radial method it is clearer to formulate the problem in the primal picture. We now solve the problem:
	\ba
	\underset{a_{\Delta}^{\mf a}}{\mbox{min/max}}\, R: \qquad \sum_{\mf a}\sum_{\Delta\in S} a_\Delta^{\mf a} \Omega(\mf a,\Delta)&=0\,,& \qquad \mbox{for all}&\quad \Omega \in \Lambda_{n_{\mbox{\tiny max}}}\\
	\sum_{\Delta \in S} a_\Delta^{\mf a} G_{\Delta}(z|\Df)&=R n^{\mf a}\,& \quad \mbox{for all}&\quad \mf a
	\ea
	Since the problem is linear in the OPE coefficients, this is still a linear program which can be solved using standard algorithms. We used the \texttt{JuliBootS} package \cite{Paulos:2014vya} here and for the computations in the next subsection.
	
\begin{figure}
		\centering
		\begin{tabular}{lr}
			\includegraphics[width=7cm]{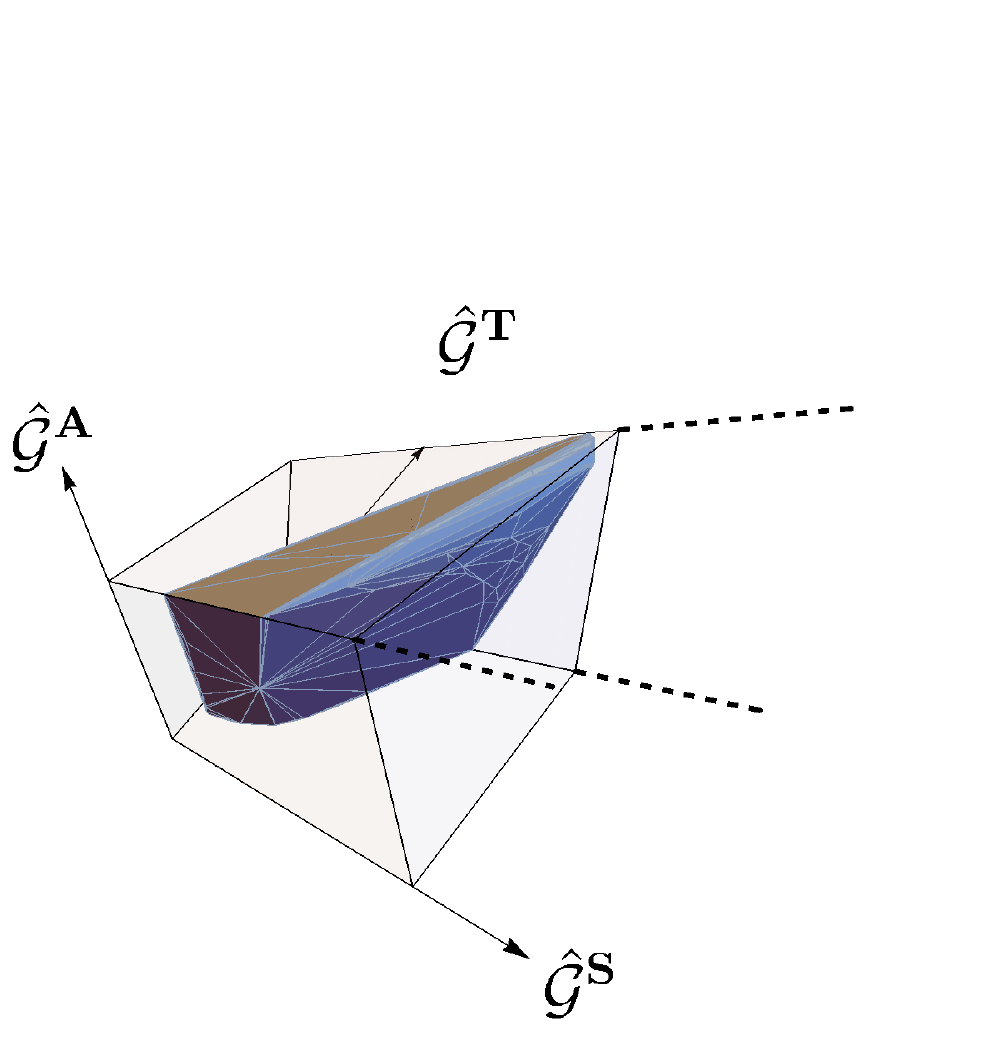}	&
			\includegraphics[width=9cm]{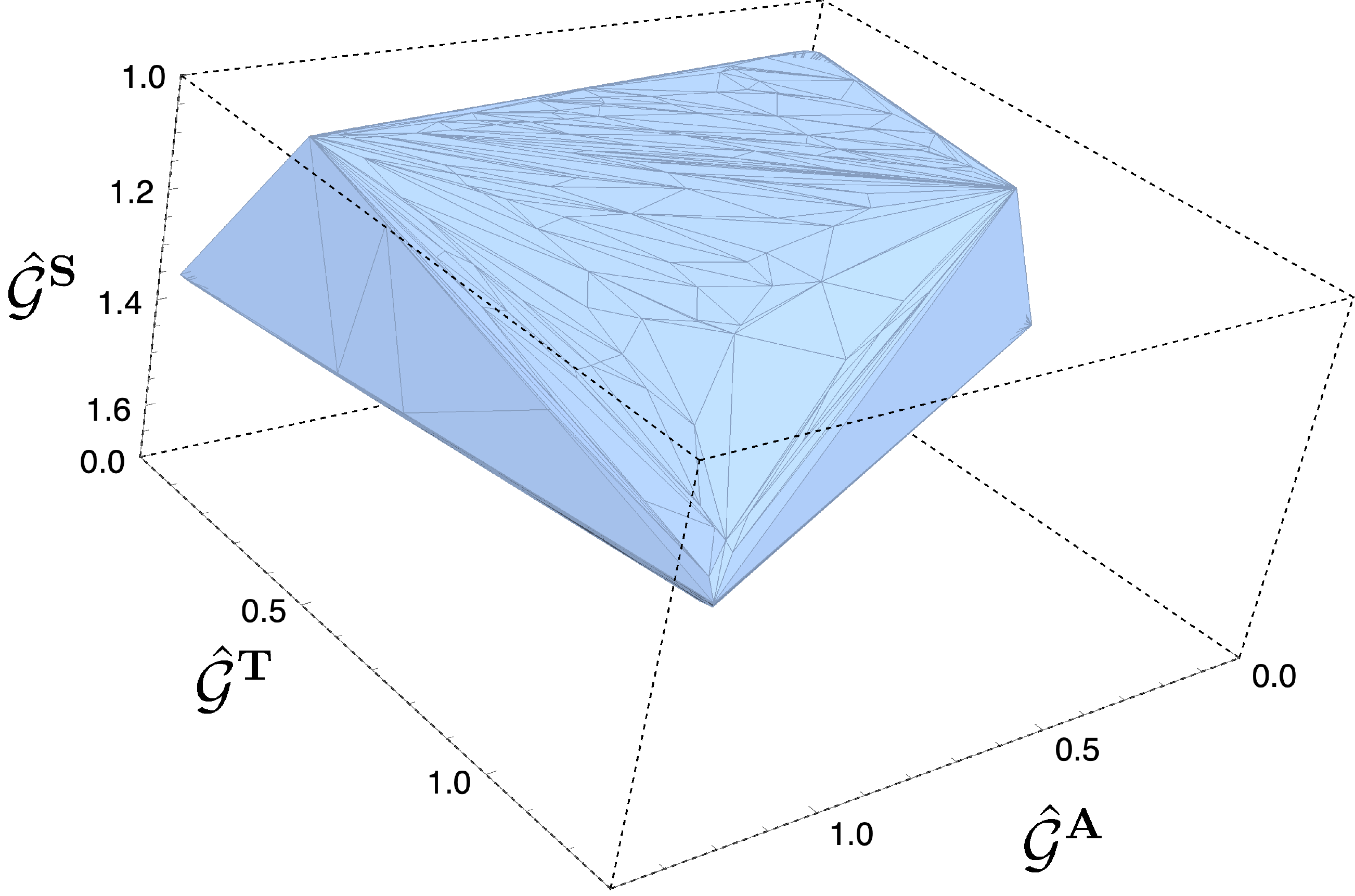}
		\end{tabular}
		\caption{\label{fig:3d}Allowed values for charged $O(N)$ correlators at $z=1/3$ with $N=2$ and with $\Df=1$. On the left we show how the dark shaded allowed region is encased in the region of positive and monotonically growing correlators. On the right the two visible triangular facets of the solid correspond to solutions where $\hat {\mathcal G}^{\mbf T}$ or $\hat{ \mathcal G}^{\mbf A}$ become constant. They meet at a corner point given by the solution $(B+F,1,1)$ discussed in the main text.
		}
	\end{figure}

	We present our numerical findings in figures \ref{fig:2d} and \ref{fig:3d}. Let us look at these in turn. Figure \ref{fig:2d} presents bounds on the correlator evaluated at the crossing symmetric point $z=1/2$. At this point the correlator values must be proportional to a $(+1)$ eigenvector of the crossing matrix. Hence the projection of the correlator onto the anticrossing symmetric eigenvector of the crossing matrix must vanish. We can use this to eliminate say $\mathcal G^{\mbf A}$:
	\ba
	\sum_{\mf a} E_{-,1}^{\mf a}\mathcal G^{\mf a}(\mbox{$\frac 12$})=0 \Leftrightarrow 
	\mathcal G^{\mbf A}(\mbox{$\frac 12$})=2\frac{\sqrt{d_{\mbf A}}}N \mathcal G^{\mbf S}(\mbox{$\frac 12$})-\frac{N+2}N\,\sqrt{\frac{d_{\mbf A}}{d_{\mbf T}}} \mathcal G^{\mbf T}(\mbox{$\frac 12$})
	\ea
	Since $\mathcal G^{\mbf A}(z)$ must be non-negative, we are left with a non-trivial constraint
	\bea
	\mathcal G^{\mbf S}(\mbox{$\frac 12$})\geq \frac{N+2}2 \frac{\mathcal G^{\mbf T}(\mbox{$\frac 12$})}{\sqrt{d_{\mbf T}}}
	\eea
	The results for different values of $N$ are qualitatively the same. Focusing on $N=2$ for definiteness we see that the convex space of correlator values is quite simple. It is to a very good approximation the convex hull of a set of six correlators. Five of these may be written explicitly. Correlators of the form $(F,F,0)$ etc represent direct sums of uncharged solutions, such that e.g.
	\ba
	(F,F,0)&\to   E_{+,1}^{\mf a} \mathcal G^F(z)\\
	(B,0,B)&\to   E_{+,2}^{\mf a} \mathcal G^B(z)
	\ea
	and so on. The $(B+F,1,1)$ solution is more peculiar. It takes the form
	\bea
	(B+F,1,1) \to \mathcal G^{\mbf S}(z)=\frac{1}{z^{2\Df}}+\frac{N}{(1-z)^{2\Df}}\,, \quad \frac{\mathcal G^{\mbf T}(z)}{\sqrt{d_{\mbf T}}}=\frac{\mathcal G^{\mbf A}(z)}{\sqrt{d_{\mbf A}}}=\frac {1}{z^{2\Df}}\,.
	\eea
	That leaves us the sixth solution, represented by a black dot. This is the correlator which minimizes $\mathcal G^{\mbf S}(1/2)$, or equivalently, that which maximizes the gap in the $\mbf S$ channel\footnote{The relation between correlator minimization and gap maximization was discussed in \cite{Paulos:2020zxx}.}. Analysing the spectrum with the extremal functional method \cite{El-Showk:2016mxr} we find it is a non-trivial interacting solution to crossing where apparently the $\mbf T$ and $\mbf A$ sectors closely match.

	Moving on, let us consider figure \ref{fig:3d}, where in this case $z=1/3$. Now the allowed region is a three-dimensional solid. This solid must be encased in the region defined by inequalities encoding positivity of the correlators and their block expansions:
	\bea
	(1-z)^{2\Df} \mathcal G^{\mf a}(1-z)\geq z^{2\Df}\mathcal G^{\mf a}(z) \geq \delta^{\mf a \mbf S}\,, \qquad z<1/2\,.
	\eea
	When $z=1/2$ these inequalities collapse the allowed region onto a plane. Note that when inequalities relating values of the correlator at different points are saturated implies that that correlator must contain only the operator of dimension zero. In the figure we can see that the allowed region is cut by two such planes, where $\mathcal G^{\mbf T}$ or $\mathcal G^{\mbf A}$ degenerate in this way. The cut surfaces then meet at a point corresponding to the $(B+F,1,1)$ solution discussed previously. It is clear that the allowed region will get substantially modified if we disallow dimension zero operators in non-singlet sectors. It would be interesting to explore this further, and also extend the analysis for different values of $N$.
	
	\subsection{The 3d Ising twist defect}
	
	Our next application is to the $Z_2$ twist defect of the 3d Ising CFT. We begin by briefly summarizing its definition and elementary properties, referring to \cite{Billo:2013jda, Gaiotto:2013nva} for further details.
	
	Consider the 3d Ising model on a cubic lattice. Now let us take a 2d semi-infinite half-plane $M$ and flip the sign of the spin-spin coupling of the Hamiltonian for all edges intersecting the plane. This defines a $Z_2$ topological domain wall. Any deformation to the plane: $M\to M'$, with the boundary line fixed, preserves the Hamiltonian if we flip all the spins between $M$ and $M'$. The boundary of $M$ is a twist line defect.  If the bulk theory flows to an IR critical point, the boundary becomes a conformal line defect. 
	
	In the continuum limit, at criticality the model without defect is a parity-invariant CFT which has the symmetry group $O^+(1,4)\times Z_2$. The twist defect breaks the bulk symmetry to $O^+(1,2)\times O'(2)$. The $O^+(1,2)$ is the spacetime symmetry group on the defect line, having the usual 1d conformal algebra $sl(2,\mathbb{R})$. It also includes the symmetry under reflection $x\to -x$ where $x$ is the defect coordinate. This is  called the $S$-parity. The global symmetry group $O'(2)$ is a double cover of rotations around the defect, with a rotation of $2\pi$ identified with the non-identity $Z_2$ transformation. 
	
	Operators on the defect can be classified according to their $S$-parity and $O'(2)$ spin. Operators that are even (odd) under $Z_2$ have (half)-integer spins. We will consider a 4-point function of real fermionic operators $\psi_i$  ($i=1,2$) which transform as spin $\frac 12$ under $O'(2)$. In the OPE of two such operators we find operators as follows \cite{Gaiotto:2013nva}:
	\be
	\psi_i \times \psi_j \supset \{\mathcal{O}^+,\mathcal{O}^-,\mathcal{S}\}\,.
	\ee
	The $\mathcal{O}^{\pm}$ operators have spin 0 and are even(odd) under $S$-parity, while $\mathcal{S}$ operators have spin-1 and are $S$-parity even. We identify them as singlet, antisymmetric and traceless symmetric irreps respectively in the $O(2)$ of which $O'(2)$ acts as a double cover. That is, the correlator
	\be
	\langle\psi_i(x_1)\psi_j(x_2)\psi_k(x_3)\psi_l(x_4)\rangle=\frac{\mathcal{G}_{ijkl}(z)}{x_{13}^{\D_\phi}x_{24}^{\D_\phi}}\,.
	\ee
	satisfies a decomposition into $O(2)$ invariant tensors. 
	
	Some of the operator dimensions in the Ising twist defect spectrum were numerically measured in Monte Carlo simulations \cite{Billo:2013jda}. In particular the dimension of $\psi_i$ was accurately determined $\D_\psi\sim 0.9187$.  The lowest dimension in the $\mathcal{O}^+$ and $\mathcal O^-$ sectors, are estimated to be $\D_s=2.27$ and $\Delta_{p_0}=2.9$ respectively. Finally, the lowest dimension in the $\mathcal S$ sector is the displacement operator $D$ with exact scaling dimension $\D_D=2$. The conformal data known in the literature and relevant to our correlator are summarized in Table \ref{numtab}.

	\begin{table}[h]
		
		\begin{center}
			\begin{tabular}{|c|c|c|c|}
				\hline
				OPE data	& Functionals & Derivatives & Monte-Carlo  \\ \hline
				$\D_\psi$ & \textit{0.9187(6)}  & \textit{0.9187(6)} & $0.9187(6)$  \\
				$\D_D$ &\textit{2} & \textit{2} &  2  \\
				$\D_s$ &  \textit{2.27(1)} & \textit{2.27} & 2.27(1) \\
				$\D_{p_0}$ & 2.841(2) & 2.92 & 2.9(2)  \\
				$a_{D}$ &  1.730(5) &1.81 &  - \\
				$a_{s}$ & 0.777(2) & 0.90  & - \\
				$a_{p_0}$ & 0.9278(2)  &0.976  & -  \\ \hline 
			\end{tabular}
			\caption{\label{numtab}Numerical estimates of OPE data in the $Z_2$ twist defect from various approaches. The operators $s$, $p_0$ and $D$ are the lowest dimension operatator in the $\mathcal O^+$, $\mathcal O^-$, $\mathcal S$ sectors, respectively $\mbf S, \mbf A$ and $\mbf T$. The rightmost column shows Monte-Carlo resuts from \cite{Billo:2013jda}, the middle one is from conformal bootstrap using a derivative expansion \cite{Gaiotto:2013nva} and the column on the left is for the present work using GFF analytic functionals. Numbers in italics were used as input.  }
		\end{center}
	\end{table}

	Here we will we search for this theory by numerically bootstrapping an $O(2)$ invariant system of correlators using the generalized free functionals. In practice this means looking for solutions to a subset of the Polyakov bootstrap equations
	\bea
	\sum_{\mf b} \sum_{\Delta} a^{\mf b} \omega(\mf b,\Delta)=0\,,\quad \mbox{for all} \quad \omega \in \Lambda_{n_{\mbox{\tiny max}}}
	\eea
	under various assumptions on the allowed states, and where the restricted set is defined by
	\ba
	\Lambda_{n_{\mbox{\tiny max}}}:=\left\{ \alpha_n^{\mf a},\beta_n^{\mf a}\,, \quad n=0,\ldots,n_{\mbox{\tiny max}}\,; \mf a=\mbf S, \mbf T,\mbf A\right\}
	\ea
	Below we have used the bosonic GFF functionals with $n_{\mbox{\tiny max}}$ ranging from 2 to 6, i.e. we have used up to 40 bootstrap equations.
	
\begin{figure}
		\begin{center}
				\includegraphics[width=12cm]{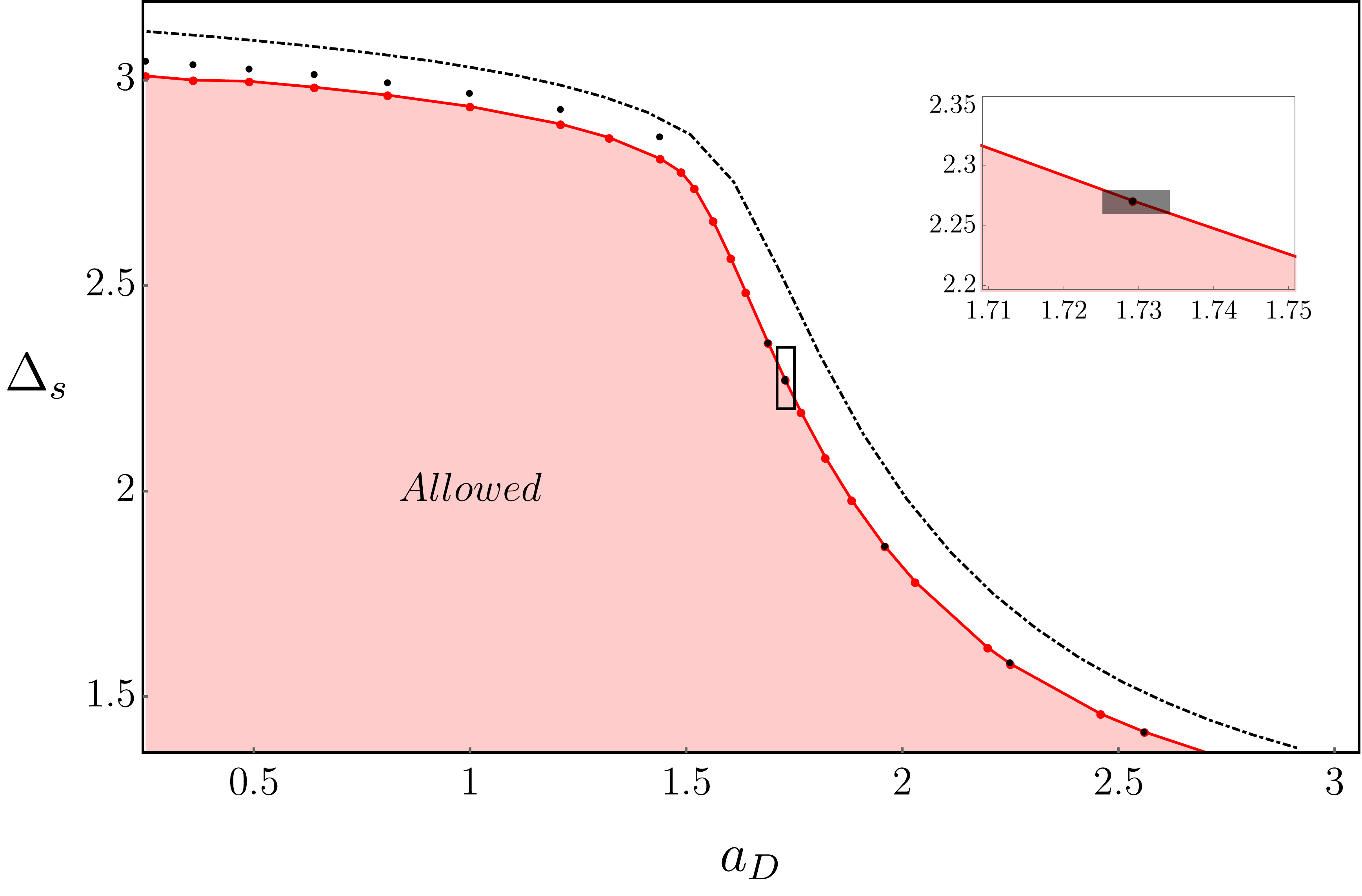}
			\caption{\label{fig:dgvsope} Maximizing the singlet gap. The red dots represent the bound obtained using GFF functionals with $n_{\mbox{\tiny max}}=6$ (40 components), black dots correspond to $n_{\mbox{\tiny max}}=2$ (16 components). The dashed-dotted black line represents the same bound using a derivative basis with 60 components \cite{Gaiotto:2013nva}. The inset shows a zoomed version of the plot around the region where the twist defect is supposed to lie, with the box representing $\Delta_s=2.27(1)$.}
		\end{center}
	\end{figure}

	We will follow the same strategy used in \cite{Gaiotto:2013nva} for the numerical bootstrap computations. We fix $\Delta_\psi=0.9187$ and assume a minimum gap of dimension $2$ in the $\mathcal{S}$ (or $\mbf T$) sector.  We then maximize the gap $\D_s$ in the singlet sector as a function of $a_D$, the OPE coefficient of the displacement operator, which sits at the gap in the $\mbf T$ sector. That is we solve the problem:\\
	\ba
	\underset{a_{\Delta}^{\mf a}\geq 0}{\mbox{max}}&\ \Delta_s \quad \mbox{s.t. for all} \quad \omega\in \Lambda_{n_{\mbox{\tiny max}}}:\\
	&\omega(0,\mbf S)+a_D\, \omega(2,\mbf T)+\sum_{\Delta\geq \Delta_s} a^{\mbf S}_\Delta \omega(\Delta,\mbf S)+\sum_{\Delta> 2} a^{\mbf T}_\Delta \omega(\Delta,\mbf T)+\sum_{\Delta\geq 0} a^{\mbf A}_\Delta \omega(\Delta,\mbf A)=0
	\ea
	The result is shown in figure \ref{fig:dgvsope}. 
	We find an important difference between using analytic functionals or derivatives. For any value of $a_{D}$ we find that bounds obtained with $n_{\mbox{\tiny max}}=2$, i.e. 16 components are better than using 60 derivatives, and in fact in a wide region we see no significant change in the bound beyond  $n_{\mbox{\tiny max}}=2$. The extremal functional method \cite{El-Showk:2016mxr} allows us to extract the spectrum of the solution saturating the bound when $\Delta_s=2.27$, which we assume is described by the twist defect. Our results are summarized in table \ref{numtab}. Again we find significant differences with the spectrum extracted using derivatives. A more general analysis of the spectrum explains why the functional basis works so well beyond the ``plateau region'', i.e. for $a_D\gtrsim 1.5$. Indeed, in that region we find that the high energy spectrum seems to quickly asymptote to that of a generalized free boson, leading to rapid convergence of the numerical bounds with the GFF basis, as explained in \cite{Paulos:2019fkw}.

	\section{Conclusions}\label{sec:conclusion}
	
	In this paper we have discussed how the analytic functional bootstrap may be extended to charged correlators. We have done this by explicitly constructing two distinct sets of analytic functionals. The existence of these two sets can be understood by the fact that there are naturally two kinds of simple extremal charged correlators, one of which is generalized free fields, and another given by a direct sum of ordinary uncharged ones. Both of these sets have an alternative formulation in terms of a Polyakov bootstrap, although the Polyakov blocks for the latter case do not have a simple AdS interpretation.
	
	In this work we have restricted ourselves to correlators of fields satisfying two basic conditions. Firstly, they should transform in a real (self-dual) irrep, and secondly the tensor product of that irrep with itself should only contain other irreps with unit multiplicity. In what concerns the first restriction, we don't foresee any particular difficulty in adapting the results of our work to general complex irreps, other than slightly modified bookkeeping. For the unit multiplicity constraint we do expect significant differences. For instance, in that case the problem of obtaining positive sum rules amounts to constructing certain positive semi-definite matrices. In this sense that problem is more similar to that of constructing functionals for a multiple correlator setup. Indeed the functionals constructed in this work can be thought of as a very useful warm-up for considering multiple correlator functionals. We hope to take this up in the near future.

	We expect the functional bases constructed in this work to be very useful for numerical applications. In this work we explored two simple applications concerning numerical bounds on scalar correlators in an $O(N)$ symmetric theory. One interesting application is to consider the large $\D_\phi$ limit, where CFT correlators become closely related to S-matrices, and we should be able to make contact with the results of\cite{Cordova:2019lot}.  In our second application, we  
	have revisited the 3d Ising twist defect and found that the new functional basis dramatically improves on the derivative basis, to the point where the numerical results seem to have essentially converged. At this point, it seems the main issue in accurate determinations of spectrum of the twist defect theory is the need to use Monte-Carlo data as an input. Hopefully a multi-correlator analysis will allow us to do away with this limitation in the near future.
	
	In theories with a global symmetry group containing a continuous parameter, such as $O(N)$, we can consider several interesting limits. One of these is $N\to \infty$. In this limit, the perturbation expansion in certain AdS theories can be resummed to obtain correlation functions with non-trivial, exact scaling dimensions in the singlet channel  \cite{Carmi:2018qzm}. In the language of this work, this would correspond to analytic functionals which are dual to bases with non-trivial spectra in the $\mbf S$ sector, and GFF spectra in the remaining ones. Given that the correlator can be exactly computed, this gives hope the same will be true for the analytic functionals. Other interesting limits correspond to sending $N\to 1$ or $N\to 0$. Both limits have the property that the crossing matrix becomes non-diagonalizable. In the $N\to 0$ limit for instance, general arguments tell us that the tensor and singlet channel spectra must collide and lead to logarithmic irreps of the conformal group. Our functional bases are in principle be perfectly adequate for studying this limit, and should be the first example of a basis dual to a logarithmic CFT. We leave a detailed analysis for further work.

	Notably absent from our work is a discussion of functionals in higher dimensions. One reason for this is that in a sense, such a basis already exists for charged correlators in higher dimensions. Indeed, the functional basis constructed in \cite{Mazac:2019shk,Caron-Huot:2020adz} can be trivially extended to the charged case, by directly acting on the crossing equations ignoring global symmetry indices. This is because that basis does not impose crossing symmetry in all three channels automatically. In contrast, extending the more recent proposal \cite{Sinha:2020win,Gopakumar:2021dvg} which is fully crossing symmetry is more non-trivial. It would be very interesting to carry this out.

\section*{Acknowledgements}
This work has benefitted from discussions with António Antunes, Lucia Cordova, Subham Dutta Chowdhury, Volker Schomerus and Aninda Sinha. The work of A.K. is supported by the German Research
Foundation DFG under Germany’s Excellence Strategy -- EXC 2121 ``Quantum Universe" -- 390833306.
We thank the Simons Collaboration on the Nonperturbative Bootstrap for leading to opportunities for discussion and collaboration.
	
	\appendix

	\section*{Appendix}
	
	\section{Contact terms}
	\label{app:contact}
	The goal of this section is to derive the form of contact terms in AdS$_2$ for both fermionic and bosonic fields, consistent with a Regge growth for large $z$ of at most a constant.
	
	\paragraph{Bosons}
	We start off with bosonic fields $\Phi_i$ in AdS$_2$, and study quartic interaction terms in the lagrangian $\mathcal L$. The Regge growth allows to have at most two derivatives. Consider first terms without any derivatives. We can write these as
	\ba
	\mathcal L \supset \sum_s g^2_{(0),s}   C^{(0),s}_{ij,kl} \, \Phi_i \Phi_j \Phi_k \Phi_l
	\ea
	where $s$ runs over the set of independent terms, the number of which we want to determine.
	Let us expand the tensor $C^{(0),s}$ into a basis of irreducible tensors,
	\ba
	C^{(0),s}_{ij,kl}=\sum_{\mf a} e^{\mf a}_s T^{\mf a}_{ij,kl}
	\ea
	where $e^{\mf a}_s$ is some vector. Now notice that Bose symmetry implies that only irreps with $\eta^{\mf a}=1$ are relevant. Hence we can choose our vectors $e^{\mf a}$ to have non-zero components only for those irreps, so apparently there seem to be $r_+$ such vectors. However, Bose symmetry also implies crossing symmetry under $i\leftrightarrow k$. Overall we must have $e^{\mf a}_s$ to be a simultaneous $+1$ eigenvector of both $\eta$ and $C$, and as we have argued in section \ref{sec:opebounds} there are $r_+-r_-$ such vectors. We conclude there are $r_+-r_-$ independent contact terms with zero derivatives.
	
	With two derivatives the interactions now can be written as
	\bea
	\mathcal L\supset \sum_s g^2_{(2),s}   C^{(2),s}_{ij,kl}\, \Phi_i \overset{\leftrightarrow}{\partial_\mu} \Phi_j \Phi_k \overset{\leftrightarrow}{\partial^\mu}\Phi_l\,.
	\eea
	We can again expand $ C^{(2),s}_{ij,kl}$ but now it will be a superposition of parity odd tensor structures, of which there are $r_-$. These are now all independent: dependence could only arise if we could construct a non-trivial tensor satisfying $C^{(2),s}_{ij,kl}=C^{(2),s}_{[ij,k]l}$, but this is impossible due to the identity $T^{\mf a}_{[ij,k]l}=0$.\footnote{Note that the quartic term in the fields has symmetries corresponding to a Young-tableau shaped like a 2x2 square, and so generically it can only be annihilated by antisymmetrization over three indices.} We conclude that in the bosonic case there are $r_+$ independent contact terms, of which $r_-$ will have two derivatives.
	
	Notice that since the terms without derivatives do not have overlap with parity odd structures, we can immediately write down a formula for the correlator describing them
	\bea
	\mathcal C_s^{(0),\mf a}(z)\propto E_{+,s}^{\mf a} \mathcal C^{(0)}(z)\,, \qquad s=1,\ldots,r_+-r_-
	\eea
	where $E_{+,s}$ is a simultaneous +1 eigenvector of $C$ and $\eta$ and $\mathcal C^{(0)}(z)$ is a contact term for the case without global symmetries.
	
	\paragraph{Fermions} For fermions the analysis is simpler since Regge boundedness now implies that we cannot include derivative terms in the lagrangian. Denoting a Majorana fermion in AdS$_2$ by $\Psi_i$, let us write contact terms as
	\bea
	\mathcal L\supset \sum_{s} \tilde g^2_{(0),s}\, \tilde {\mathcal C}^{(0),s}_{ik,jl}\, \Psi_i\cdot \Psi_j\, \Psi_k\cdot \Psi_l\,, \qquad \Psi_i\cdot \Psi_j=\epsilon^{\alpha \beta} \Psi_{i,\alpha}\Psi_{j,\beta}\,.
	\eea
	Notice we have written the indices in a slightly different manner to the bosonic case. It easy to see that $\tilde C^{(0),s}_{ij,kl}$ can be expanded in a basis of parity odd tensor structures, of which there are $r_-$. Again, these will all be independent unless we can find $\tilde C^{(0),s}_{ij,kl}=C^{(0),s}_{[ij,k]l}$, which is impossible. Hence we conclude there are $r_-$ independent Regge bounded fermionic contact terms.
	
	It turns out these can be computed explicitly. Consider a set of vectors $f_{+,s}^{\mf a}$, $s=1,\ldots,r_+$ satisfying $\eta^{\mf a} \cdot f_{+,s}^{\mf a}=f_{+,s}^{\mf a}$. Set
	\bea
	\tilde{\mathcal C}_s^{(0),\mf a}(z|\Df)=\mathcal C^{(0)}(z|\Df+1/2) \sum_{\mf b} \left[(\delta^{\mf a\mf b}-C^{\mf a\mf b})z+C^{\mf a \mf b}-\eta^{\mf a}C^{\mf a \mf b}\right] f_{+,s}^{\mf b}
	\eea
	Then it is easy to check that:
	\ba
	\tilde{\mathcal C}_s^{\mf b}(z)&=\sum_{\mf b} C^{\mf a \mf b}\tilde{\mathcal C}_s^{\mf b}(1-z)\\
	\tilde{\mathcal C}_s^{\mf a}(z)&=-(1-z)^{-2\Df} \eta^{\mf a}\tilde{\mathcal C}_s^{\mf a}(\mbox{$\frac{z}{z-1}$})
	\ea
	with the second condition guaranteeing the right OPE decomposition. Furthermore, out of these $r_+$ vectors, $r_+-r_-$ can be chosen to be +1 eigenvectors of the crossing matrix, as we recalled above, and it is easy to check that these lead to a vanishing result above. We are therefore left only with the desired set of $r_-$ independent fermionic contact terms.

	\section{GFF functional construction}
	\label{app:gffkernels}
	In this appendix we will discuss how to obtain the functional kernels of section \ref{sec:GFFfunctionals}, by showing how equation \reef{eq:fundfreeeq} can be solved systematically subject to appropriate boundary conditions.

	\subsection{Special values of $\Df$}\label{app:splval}
	We can find exact analytic solutions for special values of $\Df$. These correspond to:
	\ba
	\mbox{Fermionic GFF}\ (\epsilon=-1):&\qquad& \Df&\in \frac 12+\mathbb Z_{\geq 0}\\
	\mbox{Bosonic GFF:}\ (\epsilon=1):&\qquad& \Df&\in\mathbb Z_{\geq 0}
	\ea
	Let us set
	\ba
	p\beta^{\mf a|\mf b}_m&: \qquad& f^{\mf a|\mf b}&=\frac{2}{\pi^2} \frac{\Gamma(1+m)^2}{\Gamma(1+2m)}\left[\eta^{\mf b} \delta^{\mf a \mf b} P_{m}(\mbox{$\frac{z-2}z$})+\eta^{\mf a} C^{\mf a \mf b}\,P_{m}(\mbox{$\frac{1+z}{z-1}$}) \right]\\
	p\alpha^{\mf a|\mf b}_m&: \qquad&  f^{\mf a|\mf b}&=\frac 12 \partial_m f^{\mf b|\mf a}_{p\beta_m}-\epsilon \eta^{\mf a} C^{\mf a \mf b} \eta^{\mf b} \frac{2}{\pi^2} \frac{\Gamma(1+m)^4}{\Gamma(1+2m)\Gamma(2+2m)}\, G_{1+m}(\mbox{$\frac{1}{z}$})
	\ea
	where furthermore we require
	\bea
	\eta^{\mf a}=+1 \Rightarrow m=1,3,5,7,\ldots\\
	\eta^{\mf a}=-1 \Rightarrow m=0,2,4,6,\ldots
	\eea
	It is easy to check that these pre-functional kernels satisfy equation \reef{eq:fundfreeeq}, but in general they do not have the appropriate fall off conditions at large $z$. To construct sets of orthonormal functionals we begin with the type $\beta$:
	\bea
	\beta_m^{\mf b|\mf a}=p\beta^{\mf b|\mf a}_{\Delta^{\mf a}_0+2m-1}+\mbox{lower}
	\eea
	where we recall
	\ba
	\Delta^{\mf a}_m&=2\Df+2m\,,& \qquad \epsilon \eta^{\mf a}&=1\\
	\Delta^{\mf a}_m&=1+2\Df+2m\,,& \qquad \epsilon \eta^{\mf a}&=-1
	\ea
	Here ``lower'' stands for other terms of the form $p\beta, p\alpha$ whose index is lower (or at most, equal) than $\Delta^{\mf a}_0+2m-1$, and with coefficients chosen so that at $z=\infty$ the kernel decays faster than $1/z$. The $\alpha_m$ type kernels can then be obtained by replacing $p\beta \to p\alpha$ and differentiating those coefficients. 
	
	\subsection{General solution}
	
	Let us begin with the fundamental free equations \eqref{eq:fundfreeeq} which we restate below for convenience:
	\begin{equation}
	\label{eq:fundfreeeqApp}
	\epsilon\, \eta^{\mf a} \mathcal R_z f^{|\mf a}(z)=-(1-z)^{2\Df-2} f^{|\mf a}(\mbox{$\frac{z}{z-1}$})-z^{2\Df -2}\sum_{\mf b} \eta^{\mf a} C^{\mf b \mf a} \eta^{\mf b} f^{|\mf b}(\mbox{$\frac{z-1}{z}$})
	\end{equation}
	There are six independent structures that the kernel of a functional $\omega_0^{\mf a}$ would depend on: $1,\eta ^{\mf a}, C^{\mf a\mf b}, \eta^{\mf a}C^{\mf b\mf c}, C^{\mf a\mf b}\eta^{\mf c},\eta^{\mf a}C^{\mf b\mf c}\eta^{\mf d}$.  From the solutions  for the special values obtained in App. \ref{app:splval} we may write an ansatz for the kernels for general $\D_\phi$.  That is:
	\ba\label{ansatz}
	\tilde \omega_0^{\mf a} \ : \ \ f^{\mf a|\mf b} = \begin{cases}P^{\mf a \mf b}_+ \, h_{1,\omega}(z) \, + P^{\mf a \mf b}_- \, f_{1,\omega}(z) \, + \eta^{\mf b} C^{\mf a \mf b} \, \tilde h_{1,\omega}(z) & \text{if } \eta^{\mf a}=1\\ \\
		P^{\mf a \mf b}_+ \, h_{2,\omega}(z) \, + P^{\mf a \mf b}_- \, f_{2,\omega}(z) \, + \eta^{\mf b} C^{\mf a \mf b} \, \tilde f_{2,\omega}(z) & \text{if } \eta^{\mf a}=-1\,. \end{cases}
	\ea
	Using properties of $\eta^{\mf a}$ and $C^{\mf a\mf b}$, we may show that $h_{1,\omega}(z), \tilde h_{1,\omega}(z), h_{2,\omega}(z)$ are crossing symmetric functions (i.e. $h(z)=h(1-z)$) and $f_{1,\omega}(z),  f_{2,\omega}(z), \tilde f_{2,\omega}(z),$ are crossing antisymmetric (i.e. $f(z)=-f(1-z)$).

	We want to compute the prefunctionals as discussed in section \ref{sec:conditions}. 
	Recall the  boundary conditions from \eqref{ztoinf} and \eqref{zto0} which are:
	\be\label{ztoinfApp}
	f^{\mf a}(z)\overset{z\to \infty}{\sim} z^{-1-\varepsilon}\,.
	\ee
	\ba\label{zto0App}
	\tilde \alpha_n^{\mf a|\mf b}:&\qquad&  f^{\mf a|\mf b}(z)&\underset{z\to 0^-}\sim -\frac{2}{\pi^2}\, \frac{\eta^{\mf a}\delta^{\mf a \mf b}(\log(z)}{z^{1+\Delta^{\mf a}_n-2\Df}}\\
	\tilde \beta_n^{\mf a|\mf b}:&\qquad& f^{\mf a|\mf b}(z)&\underset{z\to 0^-}\sim \frac{2}{\pi^2}\, \frac{\eta^{\mf a}\delta^{\mf a \mf b}}{z^{1+\Delta^{\mf a}_n-2\Df}}\,.
	\ea

	One way to find the solutions is to start with a general ansatz for the functional kernels,
	\be
	F_{a,b,c,d,e}(z)=\frac{\Delta_\phi ^d \, }{w^{e+\frac{1}{2}}}  {}_3F_2\Big(c-\frac{1}{2},\frac{1}{2},2 \Delta_\phi +\frac{3}{2};\frac{a}{2}+\Delta_\phi ,\frac{b}{2}+\Delta_\phi ;-\frac{1}{4 w}\Big) \ \ \ \  \big[w=z (z-1)\big]\,,
	\ee
	with $a,b,c,d,e$ being integers.
	A good feature of the above anstaz is that it by choosing the arguments appropriately, we can get a $\log(-z)/z^k$ or $1/z^k$ ($k$ integer) behavior as $z\to 0$ as required in \eqref{zto0App}\,. Now we take a  linear combination of these functions as follows:
	\begin{align}
	h(z)&=\sum_{a,b,c,d,e} c^{h}_{a,b,c,d,e} F_{a,b,c,d,e}(z)(1-2z)\,,\nonumber\\
	f(z)&=\sum_{a,b,c,d,e} c^{f}_{a,b,c,d,e} F_{a,b,c,d,e}(z)\,.
	\end{align}
	where $h=h_{i,\omega},\tilde {h}_{1,\omega},h_{2,\omega}$ and $f=f_{1,\omega},f_{2,\omega},\tilde{f}_{2,\omega}$\,.
	We can choose suitable values of $z$ so that the equations \eqref{eq:fundfreeeqApp} give a system of simple linear equations for $c_{a,b,c,d,e}$ and then solve for these coefficients. 
	
	\paragraph{Fermionic functionals:} 
	We then proceed to find the fermionic prefunctional kernels for the simplest cases, $\tilde \alpha_0^{\mf a}$ and $\tilde \beta_0^{\mf a}$ that satisfy \eqref{zto0App}, using the above method. The actual functionals $\alpha_0^{\mf a}$ and $\beta_0^{\mf a}$ (which satisfy the large $z$ behaviour \eqref{ztoinfApp}) are obtained following subtractions shown in  \eqref{betasub}. For the global symmetry groups we can only determine the combination $ \tilde f_{2,\omega}-\frac12 f_{2,\omega}$\,, so it suffices to set  $ \tilde f_{2,\omega}=0$. The other five functions are given in the notebook included with this paper.

	
	%
	%
	%
	If one has the functionals $\a^{\mf a|\mf b}_{0}$ and $\b^{\mf a|\mf b}_{0}$, then there is a trick that allows one to get the higher functionals $\a^{\mf a|\mf b}_{n}$ and $\b^{\mf a|\mf b}_{n}$ ($n>0$). 
	This is possible through the following relation that says if $f^{|\mf b}(z)$ satisfies the free equation \eqref{eq:fundfreeeqApp} then so does (showing the $\D_\phi$ dependence):
	\be\label{trick}
	f'^{|\mf b}(z|\D_\phi)= \frac{[1+z(z-1)]^p}{[z(1-z)]^k} f^{|\mf b}(z| \D_\phi+\frac{3}{2}k-p)\,.
	\ee
	for $k,p\in \mathbb{Z}$\,.
	
	Now if we have a functional $\omega^{\mf a}_{n}$ we can obtain a shifted functional $s\omega^{\mf a}_{n+m}$ by taking $p=0$ and $k=2m$\,. So we can simply write:
	\be\label{inc-n}
	f_{s\omega_{n+m}^{\mf a}}^{|\mf b}(z|\D_\phi)= \left[\frac{1+z(z-1)}{z(z-1)}\right]^{2m} f_{\omega_{n}^{\mf a}}^{|\mf b}(z| \D_\phi+m)\,.
	\ee
	A shift in the $z$-singularity by $z^{-2m}$ results in new poles in $\D$ shifted by $2m$ in the functional action, which is exactly what we need for the GFF spectrum.  
	Given the expression for all the fermionic  functionals with the lowest $n$, it is straightforward to obtain all the shifted fermionic functionals. The functional $\omega_{n+m}^{\mf a}$ is obtained from shifted functional $s\omega_{n+m}^{\mf a}$  by subtracting lower functionals so that only the highest singularity in $z$ survives.
	
	\paragraph{Bosonic functionals:}  
 To obtain the bosonic functionals we once again use the formula \eqref{trick}. We may start with the fermionic functionals constructed above. We now need to shift the  $\D$-poles by odd number $2m+1$ compared to the fermionic values. This translates to shifting the corresponding fermionic functionals by $z^{-2m-1}$.  To obtain the components of the lowest bosonic functionals we may do the following:
	\be
	f^{|\mf b}_{s\tilde \omega^{B,\mf a}_{0}}(z)=\begin{cases}z(z-1) f^{|\mf b}_{\omega^{F,\mf a}_{0}}(z| \D_\phi-\frac 32) & \text{if } \eta^{\mf a}=1\\ \\ 
		\frac{1+z(z-1)}{z(z-1)}f^{|\mf b}_{\tilde \omega^{F,\mf a}_{0}}(z| \D_\phi+\frac 12) & \text{if } \eta^{\mf a}=-1\,. \end{cases}
	\ee
	Here $s\tilde \omega^{B,\mf a}_{0}$ denotes a shifted bosonic prefunctional, which is simply some combination of the actual prefunctionals $\tilde \omega^{B,\mf a}_{0}$. Also from the way we constructed them one should have  $s\tilde \omega^{B,\mf a}_{0}=\tilde \omega^{B,\mf a}_{0}$ for $\eta^{\mf a}=1$.
	
	The actual functionals $\omega^{B,\mf a}_{0}$ are obtained from $\tilde \omega^{B,\mf a}_{0}$ (or $s\tilde \omega^{B,\mf a}_{0}$) with the subtractions \eqref{betasub} again requiring the correct large $z$ falloff. We can obtain all higher $n$ bosonic functionals following \eqref{inc-n}.

	\section{Master functionals}
	\label{app:master}
	In this section we examine how to solve for the master functional kernels, i.e. the equation:
	\begin{multline}
	\label{eq:mastereq2}
	\epsilon\, \eta^{\mf b} \mathcal R_z f_w^{\mf a|\mf b}(z)+(1-z)^{2\Df-2} f^{\mf a|\mf b}(\mbox{$\frac{z}{z-1}$})+z^{2\Df -2}\sum_{\mf c} \eta^{\mf b} C^{\mf c \mf b} \eta^{\mf c} f^{\mf a|\mf c}(\mbox{$\frac{z-1}{z}$})\\=-\left[\delta^{\mf a,\mf b}\delta(w-z)+C^{\mf a \mf b} \delta(1-w-z)\right]
	\end{multline}
	We want to solve these equations in such a way as to remove all the homogeneous solutions, which are of course nothing but the GFF functionals. We also want for the functional to be crossing compatible, i.e. that it commutes with infinite sums of states in the crossing equation. We can achieve this by demanding
	\bea
	f^{\mf a|\mf b}(z)\underset{z\to \infty}{=} O(z^{-2})\,,\quad f^{\mf a|\mf b}(z)\underset{z\to 0^-}=\left\{
	\begin{array}{ll}
		O(\log(-z))\,, & \epsilon \eta^{\mf b}=-1 \\
		O(z^{-1})\,, & \epsilon \eta^{\mf b}=+1
	\end{array}
	\right.
	\eea
	The reason why the boundary condition at $z=0$ is different for $\epsilon\eta^{\mf b}=1$ is because in that case the $\beta_0^{\mf b}$ functional kernels don't exist (as they do not have the correct fall-off at infinity), and they would be the ones with an asymptotic behaviour $O(z^{-1})$.
	
	We can solve for the master functional kernels in two ways: by ansatz for special values of $\Df$, or numerically. We discuss these in turn.
	
	\subsection{Special values of $\Df$}
	As for ordinary functionals, we can find exact analytic solutions for special values of $\Df$. These correspond to:
	\ba
	\mbox{Fermionic GFF}\ (\epsilon=-1):&\qquad& \Df&\in \frac 12+\mathbb Z_{\geq 0}\\
	\mbox{Bosonic GFF:}\ (\epsilon=1):&\qquad& \Df&\in\mathbb Z_{\geq 0}
	\ea
	In this case we can make an ansatz for the master functional kernels:
	\bea
	f_w^{\mf a|\mf b}(z) = \begin{cases}P^{\mf a \mf b}_+ \, h_{1}(w,z) \, + P^{\mf a \mf b}_- \, f_{1}(w,z) \, + \eta^{\mf b} C^{\mf a \mf b} \, \tilde h_{1}(w,z) & \text{if } \eta^{\mf a}=1\\ \\
		P^{\mf a \mf b}_+ \, h_{2}(w,z) \, + P^{\mf a \mf b}_- \, f_{2}(w,z) \, + \eta^{\mf b} C^{\mf a \mf b} \, \tilde f_{2}(w,z) & \text{if } \eta^{\mf a}=-1\,. \end{cases}
	\eea
	where the symmetry properties are as before $h_1,h_2,\tilde h$ symmetric in $z\leftrightarrow 1-z$ and $f_1,f_2,\tilde f$ symmetric. The ansatze take the form e.g.
	\bea
	f_1(w,z)=\frac{\sum_{k=0}^M [z(z-1)]^k\left[(2z-1)c_{1,k}(w) +d_{1,k}(w) \log\left(\frac{z-1}z\right)\right]}{(z-1) z (w-z) (w (z-1)-z) (w+z-1) ((w-1) z+1)}\, 
	\eea
	The coefficient functions get uniquely fixed by \reef{eq:mastereq2} once we impose boundary conditions. As for the $\alpha,\beta$ functionals, the identity $T^{\mf a}_{[ij,k]l}=0$ implies we have the freedom to choose $\tilde f_2(w,z)=0$ in the above.

	\subsection{Dispersion relation}
	The master functional kernels satisfy two distinct dispersion relations. Since the discussion is almost identical to that of reference \cite{Paulos:2020zxx}, modulo some extra indices, we only discuss one of them here.
	
	To start with, we postulate a dispersion relation for a general set of correlation functions of the form
	\bea
	\overline{\mathcal G}^{\mf a}(w)=-\epsilon \eta^{\mf a}\int_0^1 \ud z\, \sum_{\mf b} g^{\mf a|\mf b}(w,z) \,d^2 \overline{\mathcal G}^{\mf b}(z)
	\eea
	where the bar reminds us that the above holds for suitably subtracted correlators. Our goal is to determine the appropriate kernel $g^{\mf a|\mf b}(w,z)$ and precise subtractions such that the above holds.
	
	If the dispersion relation above holds then we expect:
	\bea
	d^2_w g^{\mf a, \mf b}(w,z)=-\epsilon \eta^{\mf a}\delta^{\mf a\mf b} \delta(w-z) \label{eq:d2g}
	\eea
	This determines the discontinuities of $g^{\mf a, \mf b}$ for $z<0$ and $z>1$ as we will now see. Let us set
	\bea
	\hat g^{\mf a|\mf b}(w,z)=\frac{g^{\mf a|\mf b}(w,z)}{\sqrt{w(1-w)}}\,.
	\eea
	Then equation \reef{eq:d2g} gives:
	\bea
	\mathcal I_w \hat g^{\mf a|\mf b}(w,z)=\eta^{\mf a} w^{-2\Df-\frac 32}\, \hat g^{\mf a|\mf b}(\mbox{$\frac{w}{w-1}$},z)-\delta^{\mf a\mf b}\delta(w-\mbox{$\frac z{z-1}$})\,\frac{(1-z)^{2\Df-\frac 12}}{\sqrt{z(1-z)}}\,, \qquad w<0
	\eea
	Imposing crossing we also determine
	\bea
	\mathcal I_w \hat g^{\mf a|\mf b}(w,z)=\sum_{\mf c} \mathcal C^{\mf a\mf c} \mathcal I_w \hat g^{\mf c|\mf b}(1-w,z)\,, \qquad w>1
	\eea
	Assuming analyticity of the kernel away from the cuts at $w>1$ and $w<0$, we can use the Cauchy formula applied to $\hat g^{\mf a|\mf b}$ to express it in terms of its discontinuities. Translating the result back into $g^{\mf a|\mf b}$ we find
	\bea
	g^{\mf a|\mf b}(w,z)=K^{\mf a\mf b}(w,z)-\int_0^1 \ud w' \sum_{\mf c} K^{\mf a\mf c}(w,w') g^{\mf c|\mf b}(w',z) \label{eq:fredholm}
	\eea
	where 
	\bea
	K^{\mf a\mf b}(w,w')=\frac{(1-w')^{2\Df-\frac 12}}{\pi} \sqrt{\frac{w(1-w)}{w'(1-w')}}\,\left( \frac{\delta^{\mf a\mf b}}{w-\frac{w'}{w'-1}}+\frac{\mathcal C^{\mf a\mf b}}{w-\frac{1}{1-w'}}\right)
	\eea
	The set of equations \reef{eq:fredholm} are of Fredholm-type and can be easily solved numerically. An issue is that when $\epsilon \eta^{\mf a}=1$ there are zero mode solutions. These correspond to $\alpha_0^{\mf a}$ functional solutions which are allowed by the integral equation. In practice however this can always be subtracted out by hand since we have those functionals explicitly for any $\Df$.

\section{Functional actions from Witten diagrams}
\label{app:witten}

In this appendix, we provide details of the Witten diagrams in AdS that enter the Polyakov blocks of section \ref{sec:PolyakovBs}. The conformal block decomposition of the Polyakov blocks are related to functional actions as we described in the main text.  In our implementation of functional actions in the numerical applications of section \ref{sec:numerics} we have extensively used the results described in this appendix.		

\subsection{Witten diagram decomposition for 1d CFT}
We can decompose the exchange Witten diagrams \footnote{Recall that we add the \ $\widetilde{}$ \  to differentiate from the contact term corrected Witten diagrams entering the Polyakov Block in section  \ref{sec:PolyakovBs}.} in $s$-channel conformal blocks as shown in section \ref{sec:PolyakovBs} which we repeat for convenience \big[denoting $\widetilde{W}^{(\pm)}_{\D,\ell}=\frac 12 (\widetilde{W}^{(t)}_{\D,\ell}\pm(-1)^{\ell}\widetilde{W}^{(u)}_{\D,\ell})$ \big]:
\begin{align}\label{Wittendecom}
\widetilde{W}^{(s)}_{\Delta,\ell}(z)&=G_{\Delta}(z|\Df)+\sum_n \left[\widetilde{a}_{n,\ell}^{(s)}(\D) G_{\D_{n,\ell}}(z|\Df)+ \widetilde{b}_{n,\ell}^{(s)}(\D) \partial G_{\D_{n,\ell}}(z|\Df)\right]\,,\nonumber\\ 
\widetilde{W}^{(+)}_{\Delta,\ell}(z)&=\sum_{n}\left[ \widetilde{a}_{n,\ell}^{(t)}(\Delta) G_{\D_n^B}(z)+ \widetilde{b}_{n,\ell}^{(t)}(\Delta) \partial G_{\D_n^B}(z)\right], \\ 
\widetilde{W}^{(-)}_{\Delta,\ell}(z)&=\sum_{n}\left[ \widetilde{\bar a}_{n,\ell}^{(t)}(\Delta) G_{\D_n^F}(z)+ \widetilde{\bar b}_{n,\ell}^{(t)}(\Delta) \partial G_{\D_n^F}(z)\right]\nonumber\,,
\end{align}
for $\ell=0,1$.	 Below we chalk out how to obtain the block decomposition coefficients in a convenient way.

Following \cite{Zhou:2018sfz}  we can use the $s$- channel equation of motion (Casimir equation) to relate the exchange diagrams, $\widetilde{W}^{(s)}_{\Delta,\ell}$ to contact diagrams $\mathcal{C}^{(s)}_{\ell}$ as shown below

\begin{equation}\label{exchangecontact}
\left(\frac{1}{2}M^{AB}_{12}M_{12AB}-C_{\Delta,\ell}\right)\widetilde{W}^{(s)}_{\Delta,\ell}(z)= \sum_{\ell'=0,1}p_{\D,\ell',\ell} \mathcal{C}^{(s)}_{\ell'}(z).
\end{equation}

Here $M^{AB}_{12}$ are the conformal generators acting on the operator at $x_1$ and $x_2$ and $C_{\Delta,\ell}$ is the Casimir eigenvalue. Also $\mathcal{C}^{(s)}_{0}$ and $\mathcal{C}^{(s)}_{1}$ are the 4-point contact diagrams involving zero-derivative and two-derivative vertices respectively (see App. \ref{app:contact}) and $p_{\D,\ell',\ell}$ are some simple normalization functions (not shown explicitly).

The conformal block decomposition of the contact diagrams has the following form:
\begin{equation} \label{decomc}
\sum_{\ell'}   p_{\D,\ell,\ell} \mathcal{C}^{(s)}_{\ell'}(z)=\sum_{n} \Big[{a}_{n}^{(\mathcal{C})}(\D,\ell) G_{\D_{n,\ell}}+{b}_{n}^{(\mathcal{C})}(\D,\ell) \partial G_{\D_{n,\ell}}\Big]\,.
\end{equation}
Now acting with the equation of motion on the above equation, we can relate the coefficients $a^{(s)}_{n,\ell}$ and $b^{(s)}_{n,\ell}$ for scalars and spin $1$ exchanges as,
\begin{equation}
\begin{split} 
& \widetilde{a}^{(s)}_{n,\ell}=\frac{{a}_{n}^{(\mathcal{C})}(\D,\ell)(C_{\D_{n,\ell},\ell}-C_{\D,\ell})-b^{(\mathcal{C})}_n(\D,\ell)\partial C_{\D_{n,\ell},\ell}}{(C_{\Delta_{n,\ell}}-C_{\Delta,\ell})^2}\,,\\
& \widetilde{b}^{(s)}_{n,\ell}=\frac{{b}_{n}^{(\mathcal{C})}(\D,\ell)}{C_{\Delta_{n,\ell}}-C_{\Delta,\ell}}.
\end{split}
\end{equation}
The coefficients $a_{n}^{(\mathcal{C})}$ and $b_{n}^{(\mathcal{C})}$ are simple to evaluate (we do not show them explicitly). We present the final coefficients $\widetilde{a}^{(s)}_{n,\ell}$ and $\widetilde{b}^{(s)}_{n,\ell}$ in \cite{notebook}.

The action of $t$-channel equation of motion on $s$- channel conformal blocks is not diagonal, i.e. action of this on RHS of t-channel counterpart of \eqref{exchangecontact} is not an eigenvalue equation. But it can again be written down in terms of finite number of $s$-channel blocks as follows

\begin{equation}
{\bf{D}}^{(t)}[G_{\Delta}(z)]=\mu G_{\Delta-1}(z)+\nu G_{\Delta}(z)+\rho G_{\Delta+1}(z).
\end{equation}
where the differential operator ${\bf{D}}^{(t)}$ representing the action of Casimir equation in the $t$-channel has the following form \cite{Zhou:2018sfz},
\begin{equation}
\frac{1}{2} D_z-(\frac{1}{z}-\frac{1}{2})D_z+\Delta(\Delta-1)-(1-4\Delta_{\phi})((1-z)\frac{d}{dz})-4 \Delta_{\phi}^2 (\frac{1}{z}-\frac{1}{2})-2\Delta_{\phi}(\Delta_{\phi}-1),
\end{equation}
with $D_z=(1-z)z^2\frac{d^2}{dz^2}-z^2\frac{d}{dz}$. The coefficients $\mu,\nu,\rho$ are simple  $\D$-dependent functions. Using this we get a recursion relation for the decomposition coefficients as follows:
\begin{equation}
\rho_{n-1} b_{n-1} +\nu_n b_n +\mu_{n+1} b_{n+1}=R_n,
\end{equation}
and 
\begin{equation}
\rho_{n-1} a_{n-1}+\nu_n a_n+\mu_{n+1} a_{n+1}+\rho'_{n-1} b_{n-1}+\nu'_{n} b_n +\mu'_{n+1} b_{n+1}=S_n.
\end{equation}
Here $n\in \mathbb{Z}_{\geq 0}$,  and
\be
a_n=\begin{cases}& \widetilde{a}_{n/2 ,\ell}^{(t)} \hspace{1.6cm} \text{for even $n$} \\
	& \widetilde{\bar{a}}_{(n-1)/2,\ell}^{(t)}  \hspace{1cm} \text{for odd $n$}  \end{cases}\,, \hspace{0.5cm}
b_n=\begin{cases}& \widetilde{b}_{n/2,\ell}^{(t)} \hspace{1.6cm} \text{for even $n$} \\
	& \widetilde{\bar{b}}_{(n-1)/2,\ell}^{(t)}  \hspace{1cm} \text{for odd $n$} \, \end{cases}\,.
\ee
Given the leading $n=0$ double trace block coefficient we can use the above recursion relations to find higher double trace block coefficients.

The expressions for $R_n, S_n, \rho_n, \nu_n, \mu_n$ are presented in \cite{notebook}.

\subsection{Computing $\tilde{b}^{(t)}_{0,\ell},\tilde{a}^{(t)}_{0,\ell} $}	

To find the leading coefficients of crossed channel exchange Witten diagrams we can use the Mellin representation of the exchange Witten diagram in general dimension,
\begin{equation}
W^{(i)}_{\Delta,\ell}(u,v)=\int_{-i \infty}^{i \infty}[ds]~ [dt]~u^s v^t \Gamma^2(\Delta_\phi-s)\Gamma^2(s+t)\Gamma^2(-t) M^{(i)}_{\Delta,\ell}(s,t),
\end{equation}
for $i=s,t,u$. For the $t$-channel the Mellin amplitude can be written in the basis of continuous Hahn polynomials as follows \cite{Gopakumar:2018xqi}: \footnote{The Mellin amplitudes of contact diagrams $\mathcal C_{0}^{(i)}$ and $\mathcal C_{1}^{(i)}$ are given by order 0 and order 1 polynomials in Mellin variables.}
\begin{equation}
M^{(t)}_{\Delta,\ell}(s,t)=\sum_{\ell'} q^{(t)}_{\D, \ell |\ell'}(s) Q_{\ell', 0}^{2 s+\ell'}(t)\,,
\end{equation}
with $Q_{\ell', 0}^{2 s+\ell'}(t)$ is the continuous Hahn polynomials and $q^{(t)}_{\D,\ell|\ell'}(s)$ is given in terms of a regularized hypergeometric function defined in \cite{Ferrero:2019luz}.

For 1d CFT we have to set $u=z^2$ and $v=(1-z)^2$ and perform the $t$ integration following \cite{Ferrero:2019luz}. Then we can extract the leading decomposition coefficient of the crossed channel exchange Witten diagram  by computing  residues of double poles at $s=\Delta_{\phi}$: 
\begin{align}
\widetilde{b}_{0,\ell}^{(t)}&=\bigg[\int_{-i \infty}^{i \infty} [dt]\text{Res}\big[z^{2s}(1-z)^{2t} \Gamma^2(\Delta_\phi-s)\Gamma^2(s+t)\Gamma^2(-t) M^{(t)}_{\Delta,\ell}(s,t)\big]_{s=\D_\phi} \bigg]_{z^{2\D_\phi}\log z}\nonumber \\
&=-2q^{(t)}_{\D,\ell|0}(\D_\phi) \k_0(\D_\phi) \hspace{2cm} \Bigg[\k_{0}(s)\equiv \frac{\Gamma(s)^4}{\Gamma(2s)}\Bigg]\,.
\end{align}
From the same residue but with the $z^{2\D_\phi}$ term one gets
\be
\widetilde{a}_{0,\ell}^{(t)}=-2 \big[\partial_{\s}(q^{(t)}_{\D,\ell|0}(\s) \k_0(\s))\big]_{\s=\D_\phi} +\gamma_E 	\, \widetilde{b}_{0,\ell}^{(t)}    \,.
\ee	
The explicit expressions of $\widetilde{a}_{0,\ell}^{(t)}$ and $\widetilde{b}_{0,\ell}^{(t)}$ for $\ell=0,1$ is presented in \cite{notebook}.\footnote{The spin dependent normalization should be taken into account carefully as discussed in \cite{Ferrero:2019luz}.}

\small
\parskip=-10pt
\bibliography{mybib}
\bibliographystyle{jhep}

\end{document}